\documentclass[10pt,twocolumn,letterpaper]{article}

\usepackage{iccv}
\usepackage{times}
\usepackage{epsfig}
\usepackage{graphicx}
\usepackage{amsmath}
\usepackage{amssymb}
\usepackage{multirow}
\usepackage{verbatim}
\usepackage{subcaption}
\usepackage{diagbox}
\usepackage{booktabs}
\usepackage{xcolor}
\usepackage{color, colortbl}

\usepackage{breqn}

\definecolor{Gray}{gray}{0.9}

\usepackage[pagebackref=true,breaklinks=true,colorlinks,bookmarks=false]{hyperref}

\iccvfinalcopy % *** Uncomment this line for the final submission

\def\iccvPaperID{0000}

\ificcvfinal\fi

\begin{document}

\title{Bayesian Feature Pyramid Networks for Automatic Multi-Label Segmentation of Chest X-rays and Assessment of Cardio-Thoratic Ratio}

\author{Roman Solovyev$^{1}$\thanks{Authors contributed  equally}\\
    \and Iaroslav Melekhov$^{2}$\footnotemark[1]\\
     \and Timo Lesonen$^{3}$\thanks{Authors contributed  equally}\\
     \and Elias Vaattovaara$^{3}$\footnotemark[2] \\
     \and Osmo Tervonen$^{3,4}$ \\
     \and Aleksei Tiulpin$^{3,4}$ \\
     \and
    \small $^1$IPPM RAS, Russian Academy of Sciences, Moscow, Russia \hspace{1pt} $^2$Aalto University, Espoo, Finland\\
    \small $^3$Oulu University Hospital, Oulu Finland \hspace{1pt} $^4$University of Oulu, Oulu, Finland\\
}

\maketitle

\ificcvfinal\thispagestyle{empty}\fi

%%%%%%%%% ABSTRACT
\begin{abstract}
   Cardiothoratic ratio (CTR) estimated from chest radiographs is a marker indicative of cardiomegaly, the presence of which is in the criteria for heart failure diagnosis. Existing methods for automatic assessment of CTR are driven by Deep Learning-based segmentation. However, these techniques produce only point estimates of CTR but clinical decision making typically assumes the uncertainty. In this paper, we propose a novel method for chest X-ray segmentation and CTR assessment in an automatic manner. In contrast to the previous art, we, for the first time, propose to estimate CTR with uncertainty bounds. Our method is based on Deep Convolutional Neural Network with Feature Pyramid Network (FPN) decoder. We propose two modifications of FPN: replace the batch normalization with instance normalization and inject the dropout which allows to obtain the Monte-Carlo estimates of the segmentation maps at test time.  Finally, using the predicted segmentation mask samples, we estimate CTR with uncertainty. In our experiments we demonstrate that the proposed method generalizes well to three different test sets. Finally, we make the annotations produced by two radiologists for all our datasets publicly available.
\end{abstract}

%%%%%%%%% BODY TEXT
\begin{figure}[ht!]
    \centering
    \includegraphics[width=\linewidth]{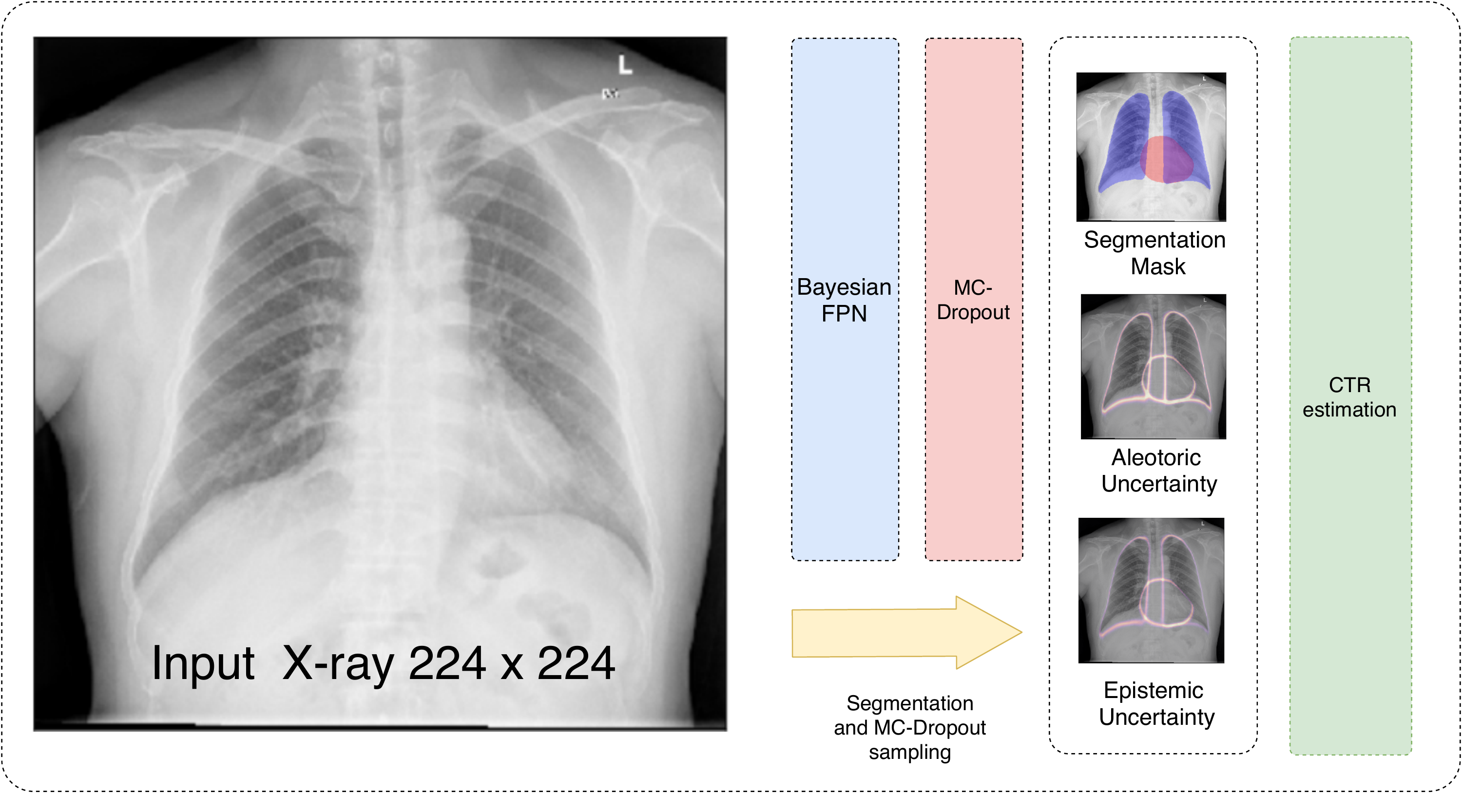}
    \caption{Overview of the workflow proposed in this study. We use our proposed modification of Bayesian FPN and perform MC-dropout inference to obtain the multilabel segmentation masks (one for the heart and one for the lungs) as well as the aleotoric and the epistemic uncertainties per image.}
    \label{fig:workflow}
\end{figure}

\section{Introduction}
Heart failure (HF) is highly prevalent in different populations. As such, its prevalence varies from 1\% to 14\% of the population according to the data from Europe and the United States~\cite{dunlay2017epidemiology}. One clinical factor having impact on the diagnosis of HF is cardiomegaly~\cite{dunlay2017epidemiology} which is a condition affecting heart enlargement. In clinical practice, assessment of cardiomegaly is trivial to a human expert (radiologist) and typically done by a visual assessment. However, there are multiple clinical scenarios when the radiologist is not available, for example in emergency care or intensive care units.

Clinically accepted quantitative measure of cardiomegaly is cardiothoratic index (CTI) -- a ratio of the heart's and the lungs' widths. In the literature, CTI is also often called a cardiothoratic ratio (CTR). CTR can be measured from chest radiographs which constitute over a half of radiographic imaging done in clinical practice~\cite{dai2018scan}.

Multiple recent studies have demonstrated promising results in assessing chest and other radiographs by applying Deep Learning (DL)~\cite{tiulpin2019multimodal,tiulpin2018automatic,wang2019grey}. These efforts indicate a possibility of reducing the amount of human effort needed for visual image assessments. Ultimately, this technology has potential to reduce the health care costs while keeping the same quality of diagnosis~\cite{saba2019present}.

DL is a methodology of learning hierarchical representations directly from data~\cite{schmidhuber2015deep}. Typically, these representations (features) are learned with respect to the task, such as image classification or segmentation. The latter allows to classify image pixels individually and eventually obtain the locations and boundaries of the objects within an image. DL-based image segmentation was shown to be a core technique in assessing CTR from chest X-rays~\cite{dong2018unsupervised,li2019automatic}. However, none of the existing CTR assessment or chest X-ray segmentation methods allow to obtain the model uncertainty which is crucial in clinical practice. %Such uncertainties can be estimated using Bayesian DL.

In this paper, we propose a robust Bayesian segmentation-based method for CTR estimation which predicts pixel-wise class labels with a measure of model uncertainty. Our approach is based on Feature Pyramid Network (FPN)~\cite{lin2017feature,seferbekov2018feature} with Resnet-50 backbone~\cite{he2016deep} and instance normalization in the decoder~\cite{ulyanov2016instance}. For uncertainty estimation, we follow~\cite{kendall2015bayesian} and utilize Monte-Carlo (MC) dropout at test time. Schematically, the proposed approach is illustrated in Fig~\ref{fig:workflow}.

The main contributions of this paper are:

\begin{enumerate}
    \item We extend traditional DL-based methods for CTR estimation to Bayesian neural network which can predict pixel-wise class labels and uncertainty bounds from segmentation masks.
    \item Compared to all the previous studies, we propose a challenging training dataset with diverse radiological findings annotated by a radiologist.
    \item The model evaluation is performed on 3 widely-used public X-ray image datasets which were re-annotated in a similar way to our training dataset, but come from different scanners and hospitals.
    \item  To the best of our knowledge, this is the first work that uses Bayesian DL in both chest X-ray segmentation and CTR estimation domains. Our methodology allows to assess the uncertainty of the model at test time, thereby providing clinical value in potential applications.
    \item Finally, we publicly release the annotations and the training dataset utilized in this study. We think that these data could set up a new, more challenging benchmark in chest X-ray segmentation.
\end{enumerate}

\section{Related Work}
\paragraph{Chest X-ray Segmentation.}
The most relevant studies to ours are by Dong~\emph{et al}~\cite{dong2018unsupervised}, by Dai~\emph{et al.}~\cite{dai2018scan} and also by Elsami~\emph{et al.}~\cite{eslami2019image}. They introduced adversarial training to enforce the consistency between the predictions and the ground truth annotations. Both studies explore the same methodology while the former one is mainly focused on CTR estimation and uses the adversarial training for unsupervised domain adaptation (UDA), the latter is rather targeting segmentation of chest X-ray. The methods demonstrate that better generalization performance to unseen data can be achieved by using adversarial training.

Besides CTR estimation realm, there are other studies approaching the segmentation problem of chest X-ray images by applying DL. Arbabshirani \emph{et al.}~\cite{arbabshirani2017accurate} and recent works~\cite{chen2018semantic,souza2019automatic} demonstrated remarkable performance in lungs segmentation. Furthermore, Wessel~\emph{et al.}~\cite{wessel2019} utilized mask R-CNN~\cite{mask_rcnn} to successfully localize, segment and label individual ribs.

From the segmentation field point of view in general, there exist multiple studies that use FPN as a decoder for image segmentation~\cite{gao2018end,kirillov2019panoptic,rakhlin2019breast}. In particular, the study by Seferbekov~\emph{et al.}~\cite{seferbekov2018feature} explores a very similar architecture to ours and seems to be the first to demonstrate a combination of ImageNet-pretrained Resnet-50 encoder with FPN decoder successfully applied to image segmentation.

In Bayesian segmentation, we note the study by Kendall~\emph{et al.}~\cite{kendall2015bayesian} that introduced the use of MC dropout for the uncertainty estimation in image segmentation. Furthermore, the recent study by~\cite{mukhoti2018evaluating} proposed to use a modification of DeepLab-v3+~\cite{chen2018encoder} that allowed to achieve state-of-the-art segmentation results and at the same time obtain uncertainty estimates.

\paragraph{Limitations of the Existing Chest X-ray Segmentation Datasets.}
In this paragraph, we also tackle an important issue of existing annotations and images in CTR assessment and Chest X-ray segmentation realm. In particular, all the existing DL-based CTR estimation aforementioned methods have been trained on the datasets that \emph{do not} include the true boundaries of the anatomical structures within the chest X-rays. While this does not have significant impact on CTR estimation in general, the absence of the true boundaries of heart and lungs limits the scope of applications that can be built using the automatic Chest X-ray segmentation (e.g. detection of plural effusion). Moreover, the existing datasets were originally from tuberculosis (TB) domain which also limits the reliable testing of the segmentation and CTR assessment models. We argue that \emph{clinically applicable} methods must be trained and tested on the datasets that have diverse radiological findings.

\section{Method}
\begin{figure*}[ht!]
    \centering
    \includegraphics[width=\textwidth]{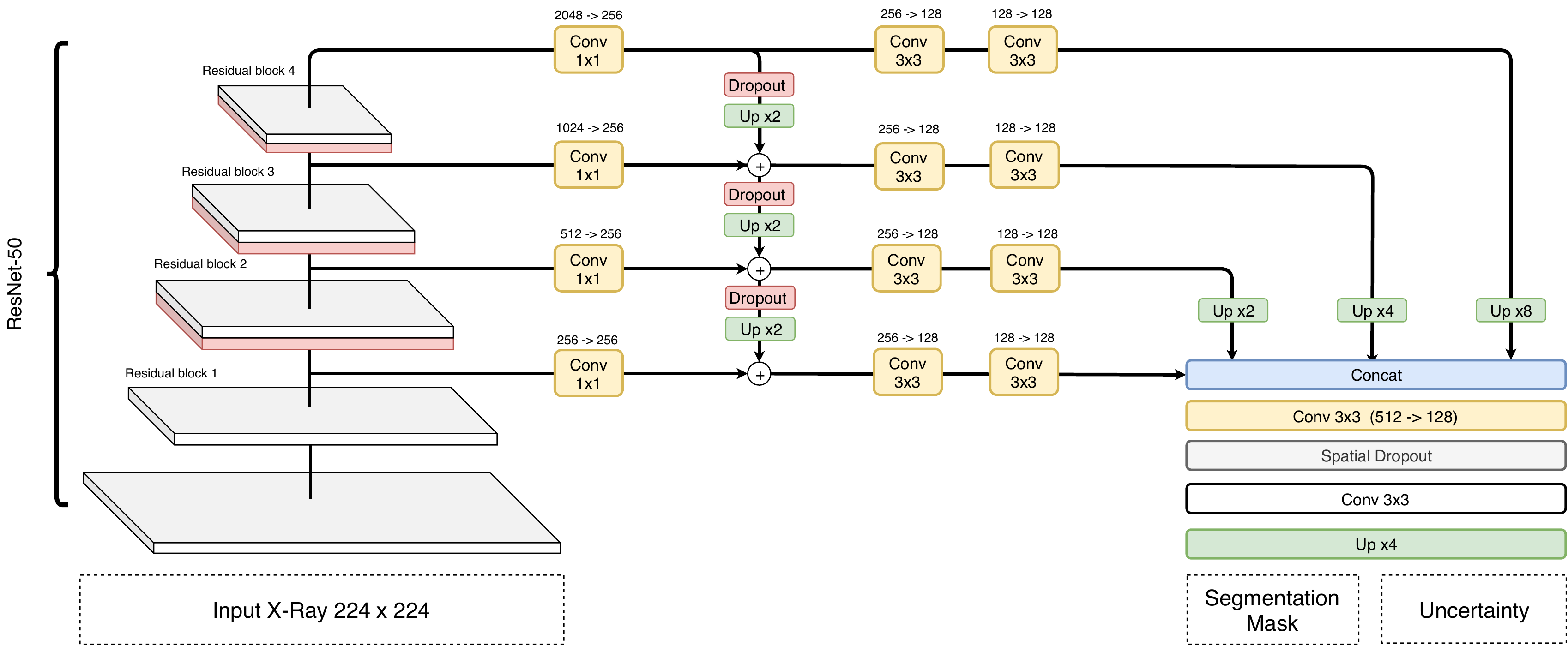}
    \caption{Model architecture. Here, we proposed a simplistic modification of FPN for image segmentation. In particular, we inserted the dropouts before the second, third and the fourth residual blocks (marked in red). Besides, we also used the dropout in the FPN part of our model and used it before every upsampling layer in the feature pyramid (nearest neighbour). However, the dropout was not used before the upsampling layers that were followed by the concatenation of the feature maps. The decoder in our model used instance normalization in the convolutional blocks (yellow)  and the final upsampling layer used a bi-linear interpolation.}
    \label{fig:arch}
\end{figure*}

\paragraph{Overview}
The proposed method is based on a combination of several state-of-the-art techniques for image segmentation. In particular, our approach leverages the power of encoder pre-trained on ImageNet~\cite{deng2009imagenet}, Feature Pyramid Networks-inspired decoder~\cite{lin2017feature,seferbekov2018feature}, instance normalization~\cite{ulyanov2016instance} and MC dropout at test time~\cite{kendall2015bayesian}. The architecture of the proposed approach is illustrated in~Fig.~\ref{fig:arch}.

\paragraph{Backbone}
We used a standard Resnet-50 pre-trained on ImageNet~\cite{deng2009imagenet}. We do not freeze the encoder during training and merely use it as is from the beginning. It is worth to note that our pre-trained encoder follows a Batch Normalization (BN) layer that learns the mean of the dataset during the training. Furthermore, we inserted the dropouts with a probability $p=0.5$ before the second, third and the fourth residual blocks of ResNet50 (see Fig.~\ref{fig:arch}).

\paragraph{Decoder}
The decoder is a standard FPN. However, similarly to Seferbekov~\emph{et al.}~\cite{seferbekov2018feature}, we do not use intermediate supervision and prediction at each layer of the feature pyramid as it is usually done for object detection~\cite{lin2017feature}. As the sizes of the feature pyramid and the input image do not match, we use nearest neighbours upsampling. In contrast to~\cite{seferbekov2018feature}, we replace each BN layer in the decoder by instance normalization (IN) layer at every level of the feature pyramid.

In addition to dropout units in the backbone, we also apply them after each $1\times 1$ convolutional block of the feature pyramid as illustrated in Fig.~\ref{fig:arch}. More specifically, the dropouts have been used in the decoder only before upsampling layers. %However, the dropout was not used before the final upsampling layers that were followed by concatenation of the feature maps.

As the task of segmenting the chest X-ray structures is multi-label, rather than multi-class, the decoder has 2 outputs, where the first plane corresponds to the heart and the second one to the lungs. Before the final output layer we used a spatial dropout with a rate of $0.1$.

\paragraph{Bayesian Segmentation Framework: Monte-Carlo Dropout.}
As mentioned previously, we leverage MC-dropout technique~\cite{gal2016dropout,kendall2015bayesian}. To capture the model's uncertainty, it is necessary to estimate the posterior distribution $p(\mathbf{W}|\mathbf{X}, \mathbf{Y})$ of the model's weights $\mathbf{W}$ given the train images $\mathbf{X}$ and the corresponding labels $\mathbf{Y}$. However, this distribution is intractable, therefore, its variational approximation $q(\mathbf{W})$~\cite{mukhoti2018evaluating}. Gal and Ghahramani~\cite{gal2016dropout} have shown that training a neural network with dropout and standard cross-entropy loss function leads to minimization of Kullback-Leiber (KL) divergence between $q(\mathbf{W})$ and $p(\mathbf{W}|\mathbf{X}, \mathbf{Y})$:
\begin{equation}
    \textrm{KL}\left(q(\mathbf{W})||p(\mathbf{W}|\mathbf{X}, \mathbf{Y})\right)\xrightarrow{}\min_{\mathbf{W}},
\end{equation}
\noindent where $q(\mathbf{W})$ is chosen to be a Bernoulli distribution.

In our experiments, we enabled the dropout layers in both encoder and FPN, respectively. We then performed the sampling of $T$ pixel-wise probability masks similarly to Kendall~\emph{et al}~\cite{kendall2015bayesian}. Here, for every pixel $\mathbf{I}(i,j)$ of the input image $\mathbf{I}$, having $T$ MC-dropout iterations, we generate the prediction $\mathbf{\hat Y}^{(t)}(i, j)$ at every $t^{th}$ MC dropout iteration and eventually  estimate $\mathbb{E}[\mathbf{\hat Y}(i, j)]$ to produce the segmentation masks $\mathbb{E}[\mathbf{\hat Y}_{h}]$ and $\mathbb{E}[\mathbf{\hat Y}_{l}]$ for the heart and the lungs, respectively.

\paragraph{Bayesian Segmentation Framework: Aleotoric and Epistemic Uncertainties.}

Besides the segmentation masks, the proposed framework also produced uncertainty estimates per pixel. As such, we computed both  \emph{aleotoric} and \emph{epistemic} uncertainties. Briefly, the former one captures the inherent noise in the data (\eg sensor noise) while the latter exhibits the model's uncertainty. Both of these uncertainties are important as aleotoric uncertainty allows to estimate the need of improving the sensor precision and the epistemic uncertainty enables to assess the need of larger training dataset~\cite{mukhoti2018evaluating}. Similarly to Mukohti and Gal~\cite{mukhoti2018evaluating}, we approximated the aleotoric uncertainty for the test examples $\mathbf{x}$ given the train data $\mathcal{D}_{train}$ as a predictive entropy $\mathbb{\hat{H}}[y|\mathbf{x}, \mathcal{D}_{train}]$:

\begin{multline}
        \mathbb{\hat{H}}[y|\mathbf{x}, \mathcal{D}_{train}] = - \sum_{c}\left(\frac{1}{T} \sum_{t}p(y=c|\mathbf{x}, \hat{w}_{t})\right) \cdot \\
        \cdot \log\left(\frac{1}{T} \sum_{t}p(y=c|\mathbf{x}, \hat{w}_{t})\right),
\end{multline}\label{eq:ent}

\noindent and the epistemic uncertainty was approximated as a mutual information $\mathbb{\hat{I}}[y, \mathbf{W} | \mathbf{x}, \mathcal{D}_{train}]$ between the predictive distribution and the posterior over the weights of the model:

\begin{multline}
    \mathbb{\hat{I}}[y, \mathbf{W} | \mathbf{x}, \mathcal{D}_{train}] = \mathbb{\hat{H}}[y|\mathbf{x}, \mathcal{D}_{train}] + \\ + \frac{1}{T}\sum_{c,t}p(y=c|\mathbf{x}, \hat{w}_{t})\log p(y=c|\mathbf{x}, \hat{w}_{t}).
\end{multline}\label{eq:mi}

In our experiments, we estimated both $\mathbb{\hat{H}}_{heart}[y|\mathbf{x},\mathcal{D}_{train}]$ and $\mathbb{\hat{H}}_{lungs}[y|\mathbf{x},\mathcal{D}_{train}]$ for each image (similarly, also $\mathbb{\hat{I}}_{heart}[y, \mathbf{W} | \mathbf{x}, \mathcal{D}_{train}]$ and $\mathbb{\hat{I}}_{lungs}[y, \mathbf{W} | \mathbf{x}, \mathcal{D}_{train}]$). For visualization purposes, we displayed the summed entropy and mutual information masks, respectively.

\begin{figure}[ht!]
    \centering
    \includegraphics[width=0.8\linewidth]{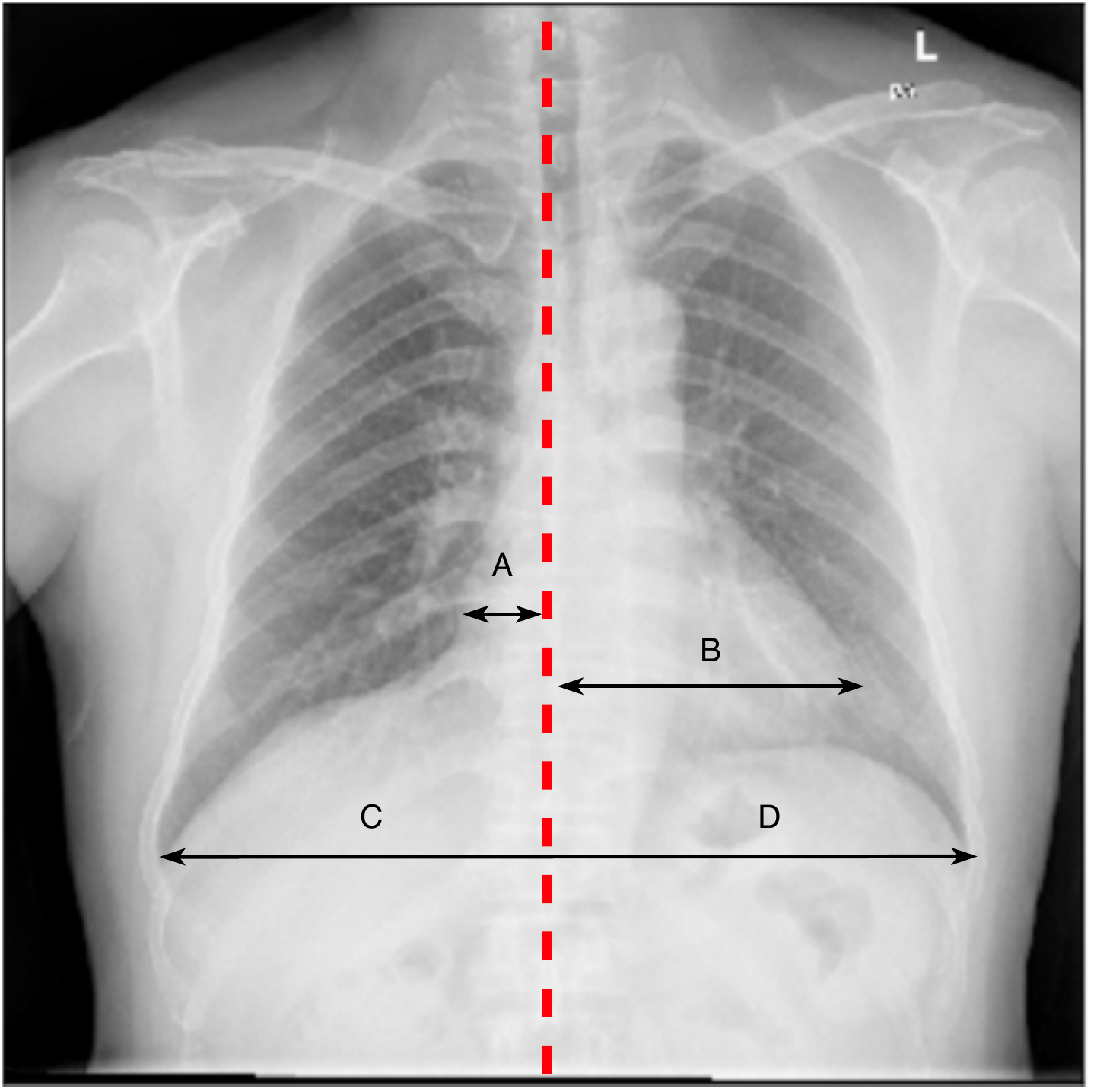}
    \caption{Visualization of the line segments used for CTR estimation.}
    \label{fig:ctr-vis}
\end{figure}

\paragraph{CTR estimation}
Following~\cite{dong2018unsupervised}, we used the same method for CTR estimation. More specifically, CTR is defined as the ratio of the widest diameter of the lungs $C+D$ and the heart $A+B$ as illustrated in Fig.~\ref{fig:ctr-vis}).

\begin{equation}
CTR=\frac{A+B}{C+D}.
\end{equation}

\paragraph{Training of the model.} During the training, we used the combination of binary cross-entropy (BCE) and soft-Jaccard loss (J) as done in various other studies~\cite{seferbekov2018feature}:

\begin{equation}
   \sum_{c={l, h}} \textrm{BCE}(\mathbf{W}, \mathbf{X}, \mathbf{Y}_c) - \textrm{J}(\mathbf{W}, \mathbf{X}, \mathbf{Y}_c)\xrightarrow{}\min_{\mathbf{W}},
\end{equation}
\noindent where $\mathbf{W}$ are the model's weights, $\mathbf{X}$ are the images and $\mathbf{Y}_h$ and $\mathbf{Y}_l$ are the ground truth masks for the heart and the lungs, respectively.

During the training process, we use various data augmentation techniques to improve the robustness and decrease possible overfitting. In particular, we used random-sized crop and $\pm 5$ degrees rotation as our main data augmentations (with a probability of $0.5$ per image). Noise addition, blur and and sharpening were used as secondary augmentations ($p=0.05$ per image). Finally, elastic distortions, Random Brightness, JPEG compression and horizontal flips were used with a probability of $0.01$ to regularize the images with hard cases.

All our models were trained with the image resolution of $224\times 224$ pixels. To train the models, we used minibatch of $8$, learning rate of $1e-4$ and spatial dropout rate of $0.1$. Our experiments were conducted in Keras~\cite{chollet2015keras} with Albumentations library~\cite{buslaev2018albumentations} used for data augmentation.

\begin{figure*}[ht!]
 	\centering
 	\begin{subfigure}[t]{.24\textwidth}
 		\centering
 		\includegraphics[width=\textwidth]{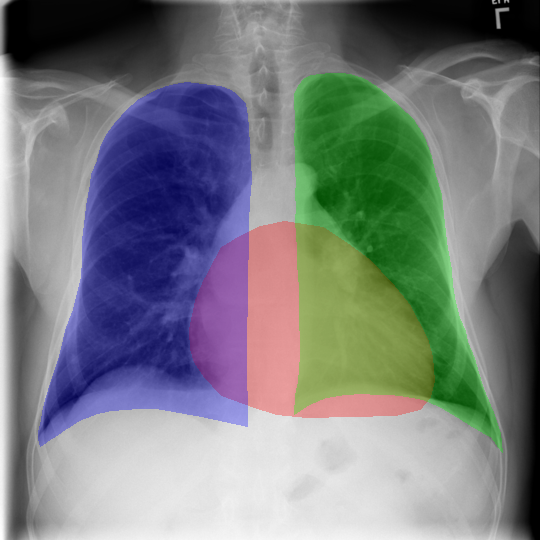}
 		\caption{Ours}
 	\end{subfigure}%
 	~
 	\begin{subfigure}[t]{.24\textwidth}
 		\centering
 		\includegraphics[width=\textwidth]{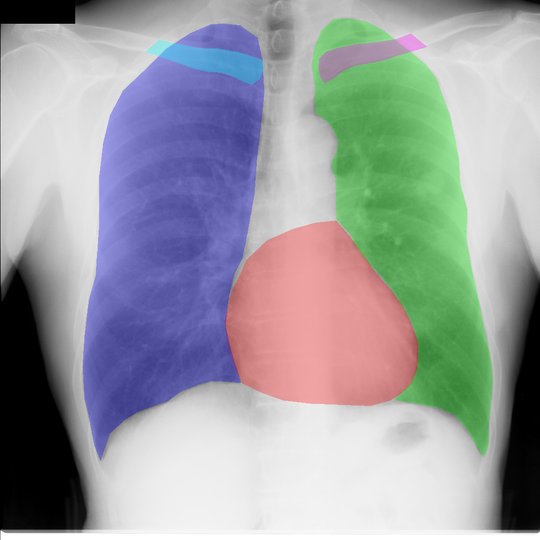}
 		\caption{JSRT}
 	\end{subfigure}%
 	~
 	\begin{subfigure}[t]{.24\textwidth}
 		\centering
 		\includegraphics[width=\textwidth]{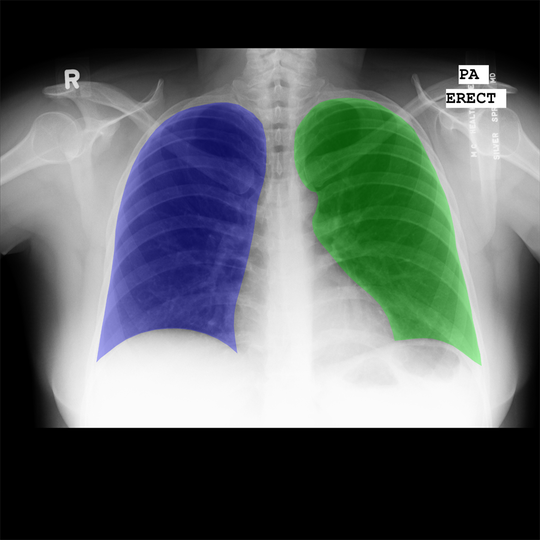}
 		\caption{Montgomery}
 	\end{subfigure}%
 	~
 	\begin{subfigure}[t]{.24\textwidth}
 		\centering
 		\includegraphics[width=\textwidth]{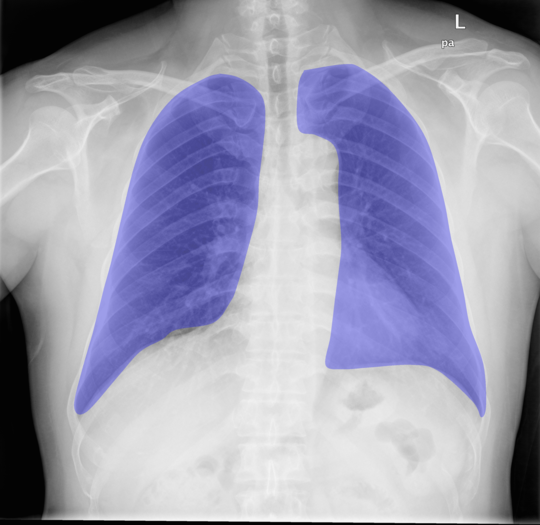}
 		\caption{Schenzen}
 	\end{subfigure}
\caption{Original annotations in all the test datasets. In our experiments we re-annotated JSRT, Montgomery and Shenzhen datasets in a similar fashion to our dataset in order to have only the image distribution different, but the segmentation masks being annotated in exactly the same fashion.}\label{fig:ann}
\end{figure*}

\begin{table*}[ht!]
    \begin{subtable}[h]{0.99\textwidth}
        \centering
        \resizebox{0.99\textwidth}{!}{%
    \begin{tabular}{l|cc|ccccc|ccc|c|cc}
\multirow{2}{*}{\diagbox[width=9em]{Decoder}{Encoder}} & vgg16 & vgg19 & ResNet18 & ResNet34 & ResNet50 & ResNet101 & ResNet152 & SE-ResNet18 & SE-ResNet34 & SE-ResNet50 & DenseNet121 & MobileNetV1 & MobileNetV2 \\
& \multicolumn{2}{c|}{\cite{Simonyan14c}} & \multicolumn{5}{c|}{\cite{he2016deep}} & \multicolumn{3}{c|}{\cite{hu2018senet}} & \cite{huang2017densely} & \cite{mobilenetv1} & \cite{Sandler2018MobileNetV2IR}  \\

\midrule
\midrule
Unet~\cite{unet} & 0.906 & \textbf{0.908} &  0.895  &  0.897 &  0.902  & 0.895 & 0.906 &  0.846 &  0.843  &  0.908 &  0.892 &  \underline{0.907}  &  0.874 \\
FPN~\cite{lin2017feature} & 0.893 & \underline{0.911} & 0.898 &  0.907  &  \underline{0.911} &  \cellcolor{Gray} \textbf{0.913} &  0.893 &  0.908 &  0.899 &  0.910 &  0.899 &  0.901 & 0.878  \\
LinkNet~\cite{linknet} & \underline{0.905} & \textbf{0.907} &  0.861 &  0.893  &  0.860  &  0.861 &  0.875  &  0.874 &  0.821 &  0.904  &  0.858  &  0.880 &  0.874  \\
PSPNet~\cite{zhao2017pspnet} & 0.871 & \textbf{0.877} &  0.852 &  0.860 &  0.859 &  0.862 &  0.865  &  0.861 &  0.859  &  \underline{0.874}  &  0.870  &  0.853  &  0.842  \\
\bottomrule
\end{tabular}
        }
    \caption{IoU \textit{Lungs}.}\label{subtab:iou_lungs}
    \end{subtable}

    \begin{subtable}[h]{0.99\textwidth}
        \centering
        \resizebox{0.99\textwidth}{!}{%
    \begin{tabular}{l|cc|ccccc|ccc|c|cc}
\multirow{2}{*}{\diagbox[width=9em]{Decoder}{Encoder}} & vgg16 & vgg19 & ResNet18 & ResNet34 & ResNet50 & ResNet101 & ResNet152 & SE-ResNet18 & SE-ResNet34 & SE-ResNet50 & DenseNet121 & MobileNetV1 & MobileNetV2 \\
& \multicolumn{2}{c|}{\cite{Simonyan14c}} & \multicolumn{5}{c|}{\cite{he2016deep}} & \multicolumn{3}{c|}{\cite{hu2018senet}} & \cite{huang2017densely} & \cite{mobilenetv1} & \cite{Sandler2018MobileNetV2IR}  \\
\midrule
\midrule
Unet~\cite{unet} & \underline{0.843}  & 0.838  &  0.805  &  0.731 &  0.822  &  0.805 &  0.820 &  0.714 &  0.750 &  \textbf{0.848}  &  0.793 &  0.791 &  0.750  \\
FPN~\cite{lin2017feature} & 0.814  & 0.786  & 0.799   &  0.836  &  \cellcolor{Gray} \textbf{0.865}   & \underline{0.863}  &  0.806   &  0.819  &  0.806   &  0.814  &  0.849 &  0.787 & 0.766 \\
LinkNet~\cite{linknet} & \underline{0.834}  & 0.808 &  0.797 &  0.766  &  0.773 &  0.799  &  0.814  &  0.762 &  0.668 &  \textbf{0.839} &  0.687 &  0.755  &  0.734  \\
PSPNet~\cite{zhao2017pspnet} & 0.764 & \textbf{0.814} &  0.740 &  0.776 &  0.717 &  0.712  &  0.745  &  0.703 &  0.741 &  0.779 &  \underline{0.781} &  0.654  &  0.661 \\
\bottomrule
\end{tabular}
        }
    \caption{IoU \textit{Heart}}\label{subtab:iou_heart}
    \end{subtable}
\caption{\textbf{Ablation study}. IoU metric (higher is better) computed on the proposed dataset for Lungs (Tab.~\ref{subtab:iou_lungs}) and Heart (Tab.~\ref{subtab:iou_heart}) obtained by different encoder-decoder architectures. For each decoder we indicate the best model as \textbf{bold} and the second best model as \underline{underscore}. The best combination of Encoder+Decoder is highlighted as \colorbox{gray!10}{\textbf{gray}}. We chose ResNet50 model as a backbone network and FPN decoder for further experiments. See Sec.~\ref{ssec:ablation_study} for more details.}\label{tab:ablation}
\end{table*}

\begin{table}[ht!]
    \begin{subtable}[h]{0.47\textwidth}
        \centering
        \resizebox{\textwidth}{!}{%
    \begin{tabular}{l|ccc}
\multirow{2}{*}{\diagbox[width=12em]{Decoder}{Normalization}} & batch (BN) & group (GN) & instance (IN) \\
 & \cite{batch_norm} & \cite{group_norm} & \cite{ulyanov2016instance} \\

\midrule
\midrule
Unet~\cite{unet} & 0.902 & 0.886  &  0.911  \\
LinkNet~\cite{linknet} & 0.860 & 0.910 &  0.915 \\
FPN~\cite{lin2017feature} & \textbf{0.911} & \textbf{0.914} & \textbf{0.916} \\
\bottomrule
\end{tabular}
        }
    \caption{Lungs.}\label{tab:lungs-normalization}
    \end{subtable}

    \begin{subtable}[h]{0.47\textwidth}
        \centering
        \resizebox{\textwidth}{!}{%
    \begin{tabular}{l|ccc}
\multirow{2}{*}{\diagbox[width=12em]{Decoder}{Normalization}} & batch (BN) & group (GN) & instance (IN) \\
 & \cite{batch_norm} & \cite{group_norm} & \cite{ulyanov2016instance} \\
\midrule
\midrule
Unet~\cite{unet} & 0.822  & 0.822  &  0.870   \\
LinkNet~\cite{linknet} & 0.773  & 0.828  &  0.862 \\
FPN~\cite{lin2017feature} & \textbf{0.865} & \textbf{0.868} & \textbf{0.884} \\
\bottomrule
\end{tabular}
        }
    \caption{Heart.}\label{tab:heart-normalization}
    \end{subtable}
\caption{Selection comparison of different normalization techniques in decoders. Here, we did not experiment with PSPNet~\cite{zhao2017pspnet} decoder due to its low performance in backbone selection stage.}\label{tab:ablation-normalization}
\end{table}

\section{Experiments and Results}
\subsection{Dataset}
Our training set was derived from ChestXray14 dataset~\cite{Wang_2017_CVPR}. The original dataset included $112,120$ chest radiographs from $30,805$ patients, while the train data used in this study comprised $421$ images randomly sampled from these data. For training, we used $294$ images images from $85$ distinct patients. $38$ images from $12$ patients were used for validation and the remaining $99$ images from $23$ patients were eventually used as a test set.

The proposed multi-label dataset has the following findings and labels (the number of samples for each label is given in parenthesis): Cardiomegaly ($27$), Emphysema ($30$), Effusion ($76$), Hernia ($9$), Infiltration ($72$), Nodule ($18$), Atelectasis $(51$), Mass ($37$), Pneumothorax ($34$), Pneumonia ($3$), Pleural thickening ($13$), Fibrosis ($14$), Consolidation ($15$), Edema ($3$). There are $177$ samples with no findings. Such label distribution makes our data more challenging for segmentation compared to the previous studies.

It is worth to mention that the original data were provided with the labels mined from the radiology reports, however the dataset did not have any segmentation masks. Our radiologist (radiologist A) annotated the train, validation and the test sets. An example of the annotations is presented in Fig.~\ref{fig:ann}. Compared to the other existing datasets illustrated in the same figure, our annotations delineate the true anatomical contours which makes the segmentation more challenging.

\subsection{Auxiliary Test Datasets}
\paragraph{Re-annotation of the existing datasets}
We evaluated our method on three auxiliary test sets each derived from three \emph{independent public datasets}, respectively (see below). The original annotations for these datasets did not include the true boundaries of the lungs underlying the heart, or had missing annotations of the heart. To evaluate the performance of our method trained on our datasets with true lung boundaries, practicing radiologist B having the same experience to the radiologist A, annotated $50$ random images for the test evaluation from each of the auxiliary test datasets. The comparison between the original annotations in the test datasets and our annotations is presented in Fig.~\ref{fig:ann}.

\paragraph{JSRT.}
Japanese Society of Radiological Technology dataset was first released in~\cite{shiraishi2000development} and complemented by the annotations from~\cite{van2002automatic}. The dataset contains 247 radiographs in total, having 154 radiographs with and 93 radiographs without lung nodules, respectively.
\paragraph{Montgomery County X-ray Set.}
This dataset contains $138$ chest radiographs acquired by the Department of Health and Human Services, Montgomery County, Maryland, USA~\cite{jaeger2014two}. Out of the total amount, $58$ subjects had and $80$ subjects did not have tuberculosis (TB).

\paragraph{Shenzhen Hospital X-ray Set.}
Shenzhen dataset is similar to Montgomery dataset as it includes the data from subjects with and without TB. These data were acquired at Shenzhen No.3 Hospital in Shenzhen, Guangdong providence, China. The total dataset size was of $340$ images having $275$ images from patients with TB~\cite{jaeger2014two}.

\begin{figure}[t!]
 	\begin{subfigure}[t]{.48\linewidth}
 		\centering
 		\includegraphics[width=\textwidth]{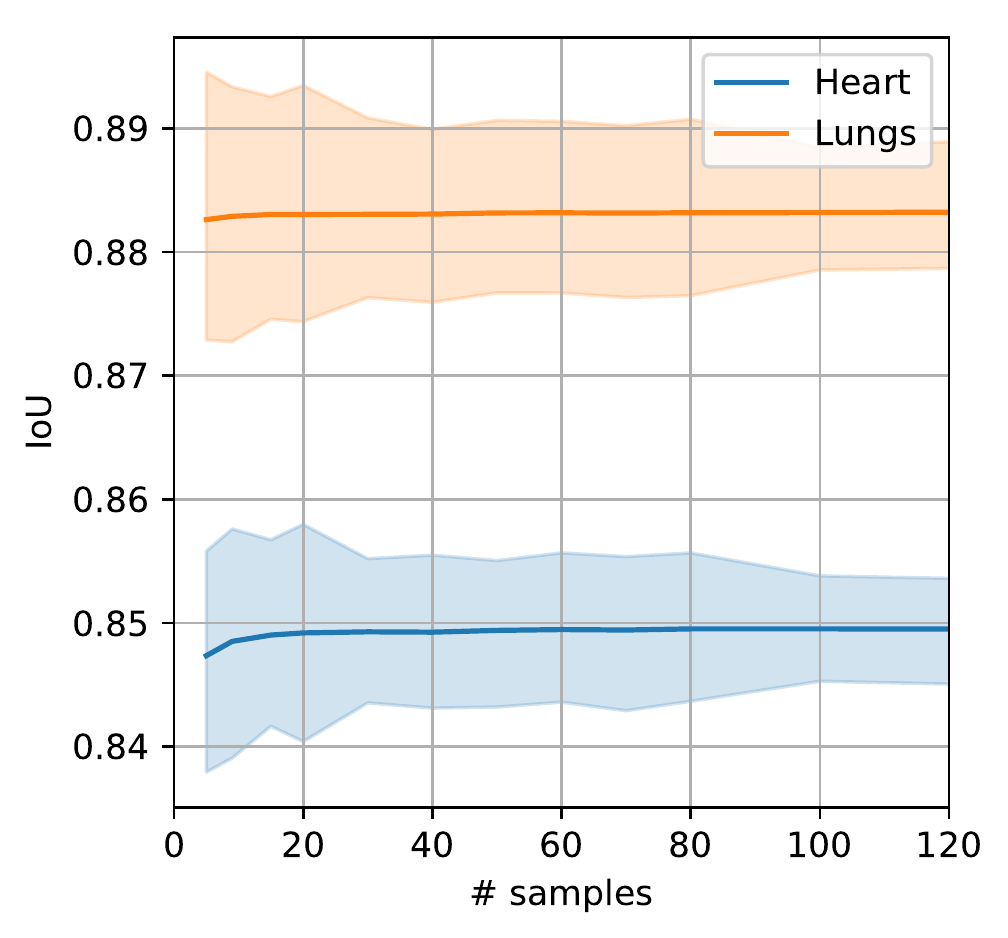}
 		\caption{IoU per organ.}
 	\end{subfigure}%
 	~
 	\begin{subfigure}[t]{.48\linewidth}
 		\centering
 		\includegraphics[width=\textwidth]{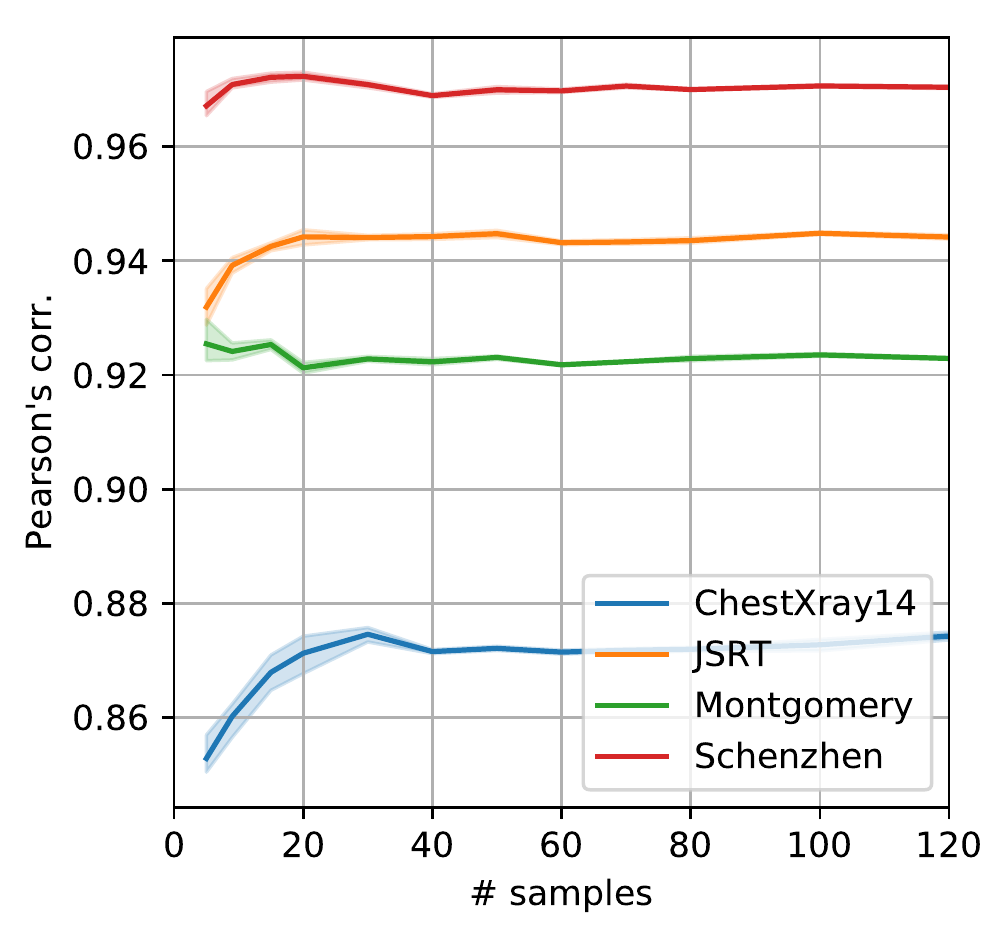}
 		\caption{CTR correlations with the ground truth per dataset.}
 	\end{subfigure}
 	\caption{Graphical illustration of dependency between the performance on the test sets and the number of MC dropout samples. 95\% intervals are also highlighted and very computed using bootstrapping.}\label{fig:mc-iterations-qualitative}
\end{figure}

\subsection{Ablation study}\label{ssec:ablation_study}
\paragraph{Overview}
This section describes the ablation study conducted on deterministic models. After selecting the best deterministic model, we inserted the dropout as shown in Fig.~\ref{fig:arch}. We first investigated what combination of encoder and decoder is relevant to our task and, subsequently, analyzed different types of normalization in the decoder.

\paragraph{Backbone and Decoder}
Latest advances in image segmentation demonstrate that transfer learning is helpful in image segmentation. As such, the use of encoders pre-trained on ImageNet~\cite{deng2009imagenet} is a core technique in all the current state-of-the-art segmentation networks~\cite{kirillov2019panoptic}. In our study, we investigated multiple pre-trained models with state-of-the-art decoders, namely U-Net~\cite{ronneberger2015u}, FPN~\cite{lin2017feature}, LinkNet~\cite{linknet} and PSPNet~\cite{zhao2017pspnet} (Tab.~\ref{tab:ablation}).

Our experiments demonstrate that in lung segmentation, ResNet101 and ResNet50  with FPN decoder yielded the best and second best results, respectively (Tab.~\ref{subtab:iou_lungs}). In heart segmentation, Resnet50 backbone outperformed Resnet101 and both of the encoders with FPN decoder achieved best and second best results, respectively. Therefore, for simplicity and speed of computations, we selected Resnet50 with FPN to be our main configuration.

\paragraph{Normalization in the Decoder}
Once the best configuration has obtained, we assessed the influence of normalization in the decoder. In particular, we investigated whether replacement of batch normalization to group or instance normalization has any effect on the performance of our model. These experiments are presented in Tab.~\ref{tab:ablation-normalization}. The results demonstrate that instance normalization achieves better performance compared to other normalization types.

\begin{table}[t!]
\footnotesize
\centering
\begin{tabular}{@{}lllllc@{}}
\toprule
\multicolumn{1}{c}{\multirow{2}{*}{\textbf{Dataset}}} & \multicolumn{2}{c}{\textbf{Heart}} & \multicolumn{2}{c}{\textbf{Lungs}} & \multicolumn{1}{c}{\multirow{2}{*}{\textbf{Pearson's corr.}}} \\ \cmidrule(lr){2-5}
\multicolumn{1}{c}{} & IoU & \multicolumn{1}{l}{Dice} & IoU & \multicolumn{1}{l}{Dice} & \multicolumn{1}{c}{} \\ \midrule
ChestXray14 & 0.87 & 0.93 & 0.92 & 0.96 & 0.87 \\
JSRT & 0.82 & 0.90 & 0.87 & 0.93 & 0.95 \\
Schenzhen & 0.84 & 0.91 & 0.87 & 0.93 & 0.97 \\
Montgomery & 0.86 & 0.92 & 0.87 & 0.93 & 0.92 \\ \bottomrule
\end{tabular}
\caption{Quantitative results for each of the datasets. Here, we present the IoU for the Lungs and the Heart and also the Pearson's correlation between the ground truth CTR (computed from the manually annotated masks) and the predicted CTR (computed from the predicted segmentation masks).}
\label{tab:quantitative-results}
\end{table}

\subsection{Test Set Results}
\paragraph{Optimal Number of Monte-Carlo Samples}
In our experiments, we assessed the influence of MC dropout onto the performance of our segmentation and CTR estimation pipeline. As such, we computed the aggregated IoU  values for heart and lungs ground truth masks. Besides, we also computed the Pearson's correlation of CTR computed using the ground truth and the predicted masks for different number of MC samples. These results are visualized in Fig.~\ref{fig:mc-iterations-qualitative}. From this plot it can be seen that the optimal number of iterations on all datasets with respect to IoU and CTR correlations is close to $20$. We use this number for our further evaluation of the developed method.

\paragraph{Quantitative Results}
For the optimal number of MC samples (20 according to~Fig.~\ref{fig:mc-iterations-qualitative}), we  performed the quantitative evaluation of our model on all the test datasets, namely ChestXray14, JSRT, Shenzhen and Montgomery. Besides computing the IoU, similarly to the previous paragraph, we also computed the Pearson's correlation between the CTR values for manually annotated masks and the predictions. These results are presented in Tab.~\ref{tab:quantitative-results}.

\begin{figure*}[ht!]
 	\centering
 	\begin{subfigure}[t]{\textwidth}
 		\centering
 		\includegraphics[width=.24\textwidth]{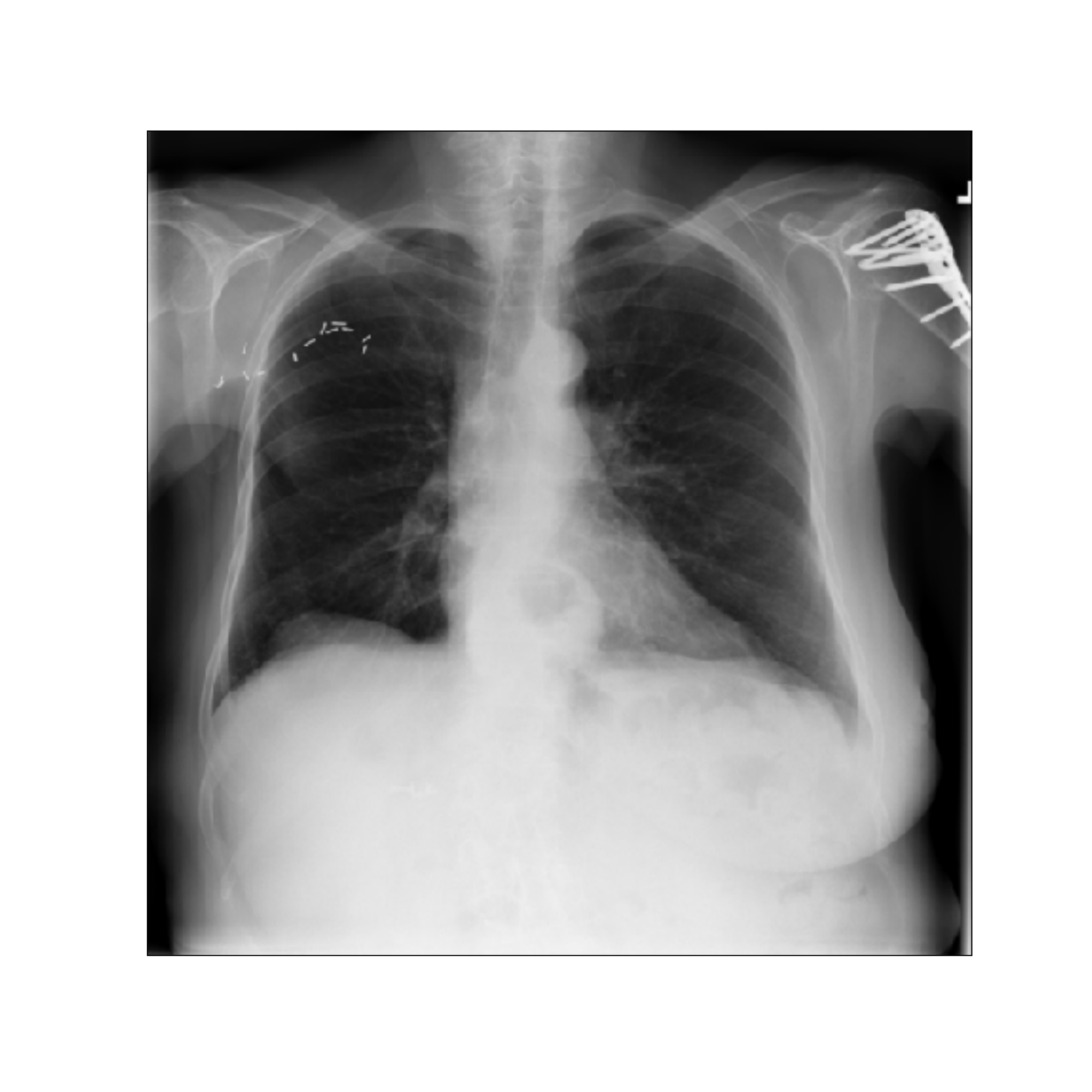}
 		\includegraphics[width=.24\textwidth]{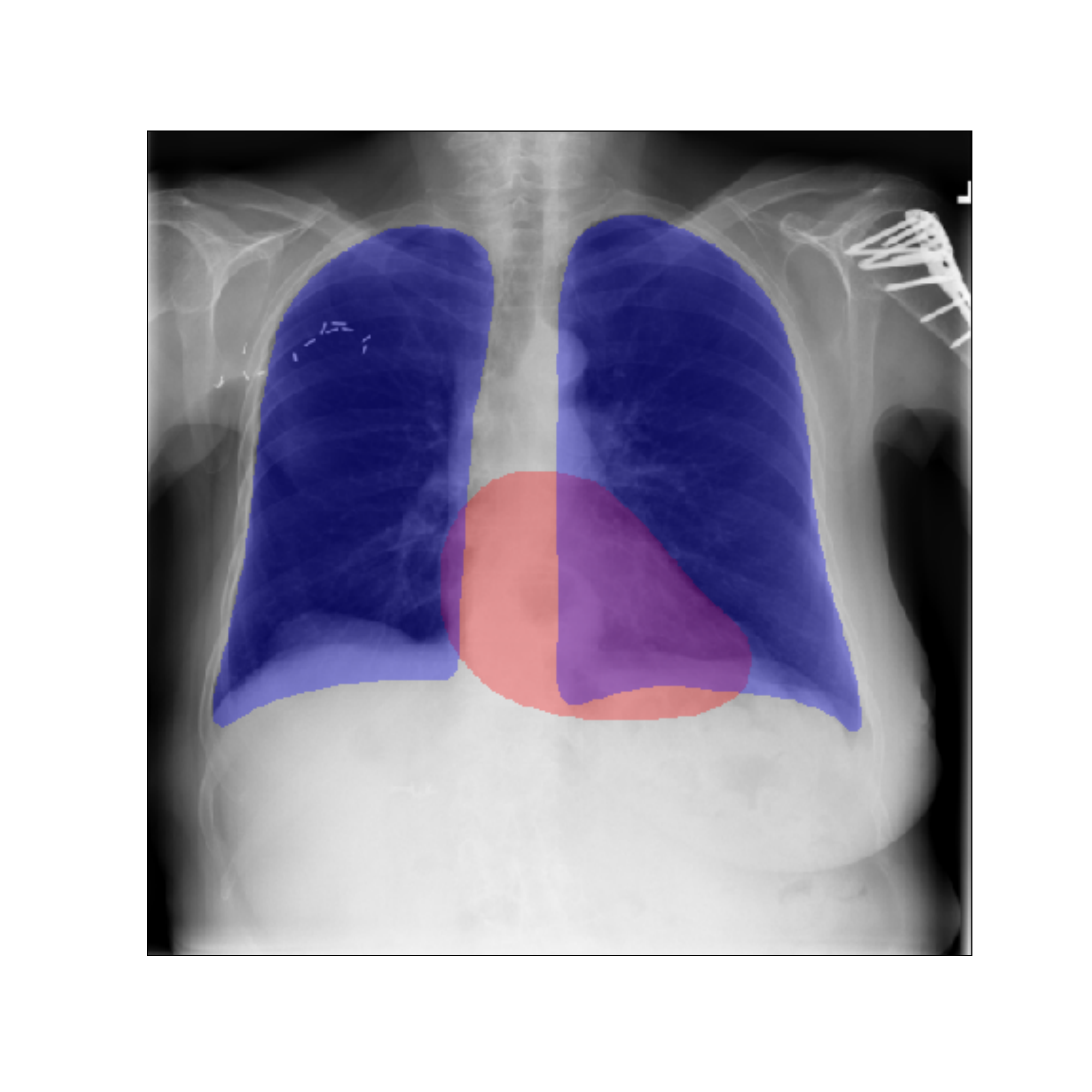}
 		\includegraphics[width=.24\textwidth]{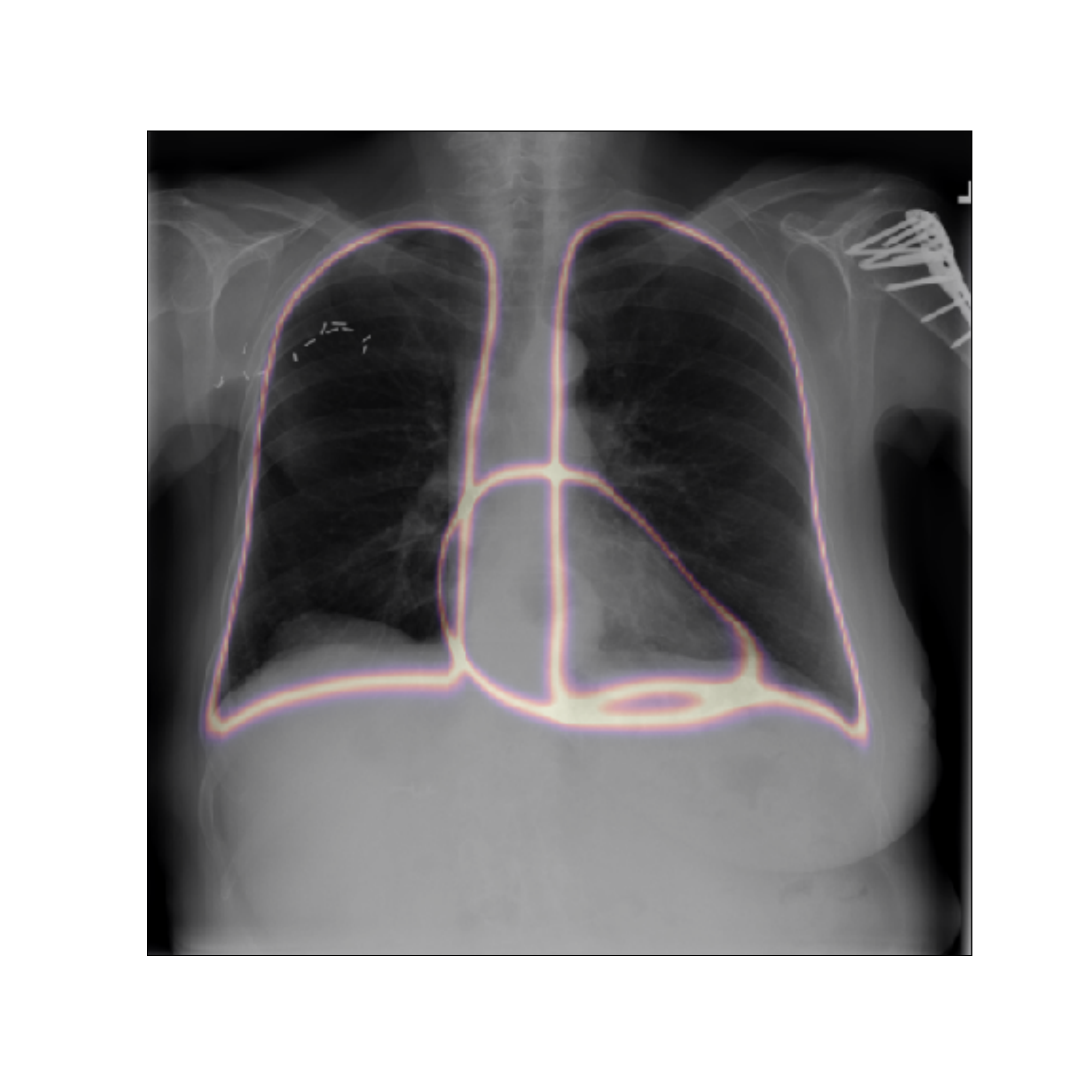}
 		\includegraphics[width=.24\textwidth]{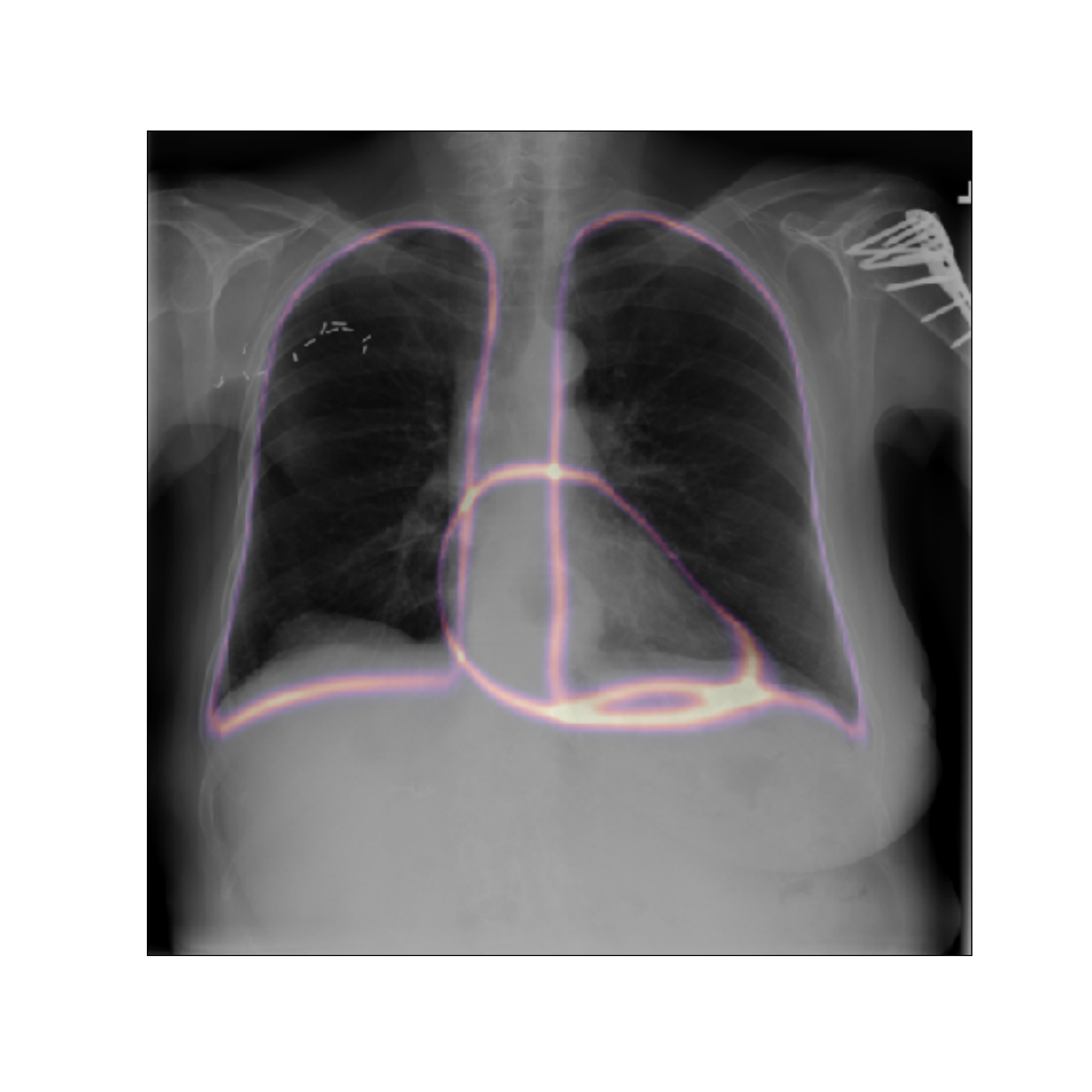}
 		\caption{ChestXray14}
 	\end{subfigure}

 	\begin{subfigure}[t]{\textwidth}
 		\centering
 		\includegraphics[width=.24\textwidth]{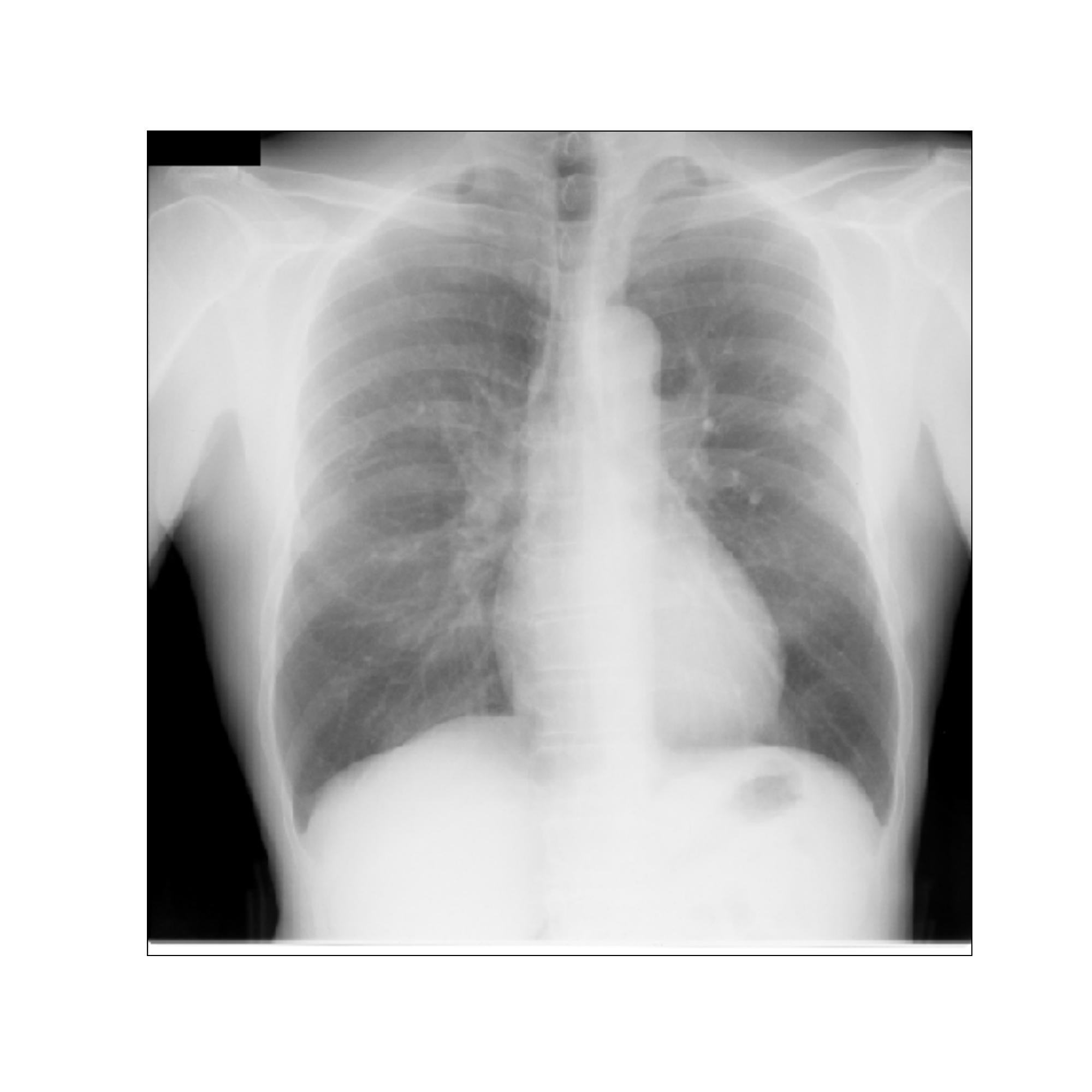}
 		\includegraphics[width=.24\textwidth]{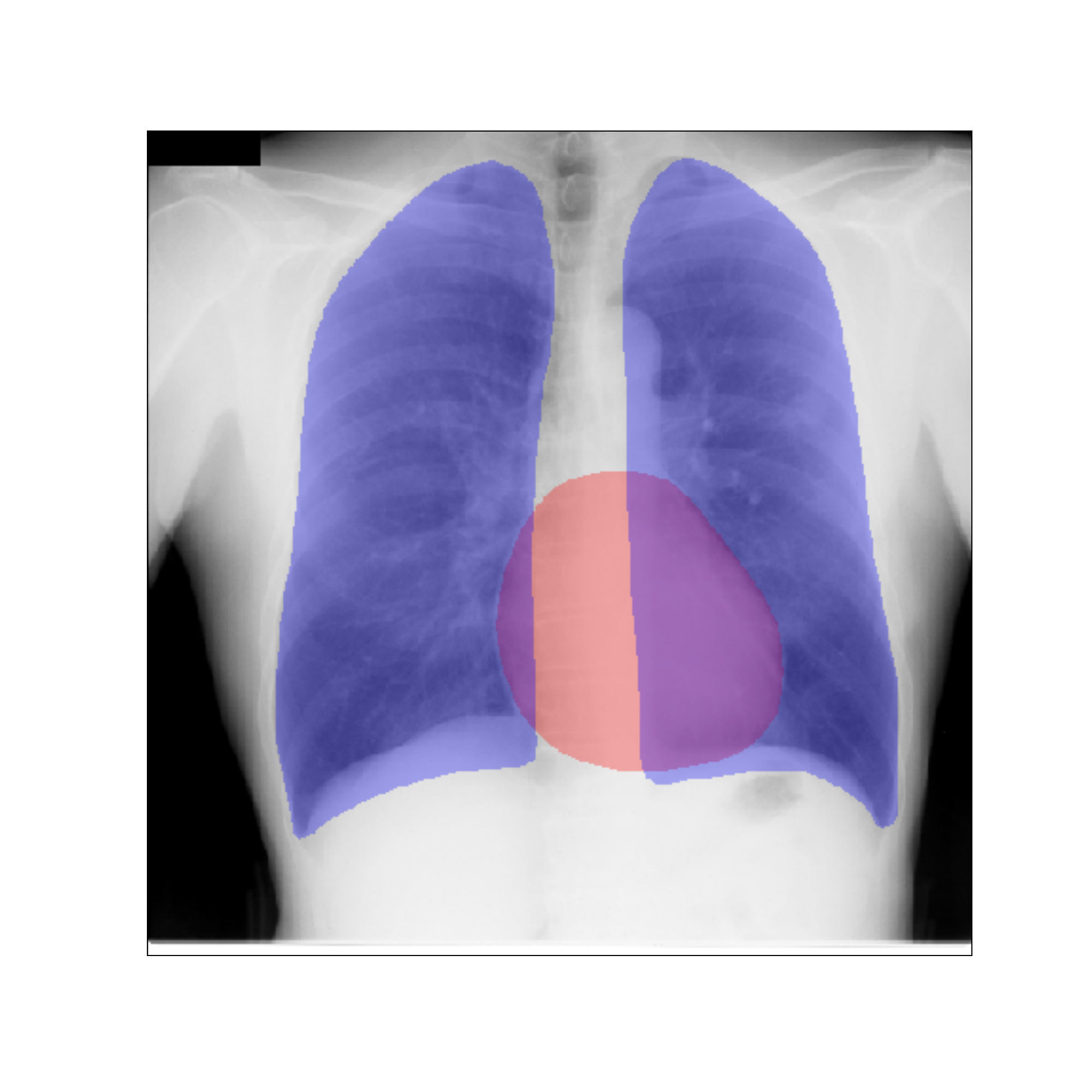}
 		\includegraphics[width=.24\textwidth]{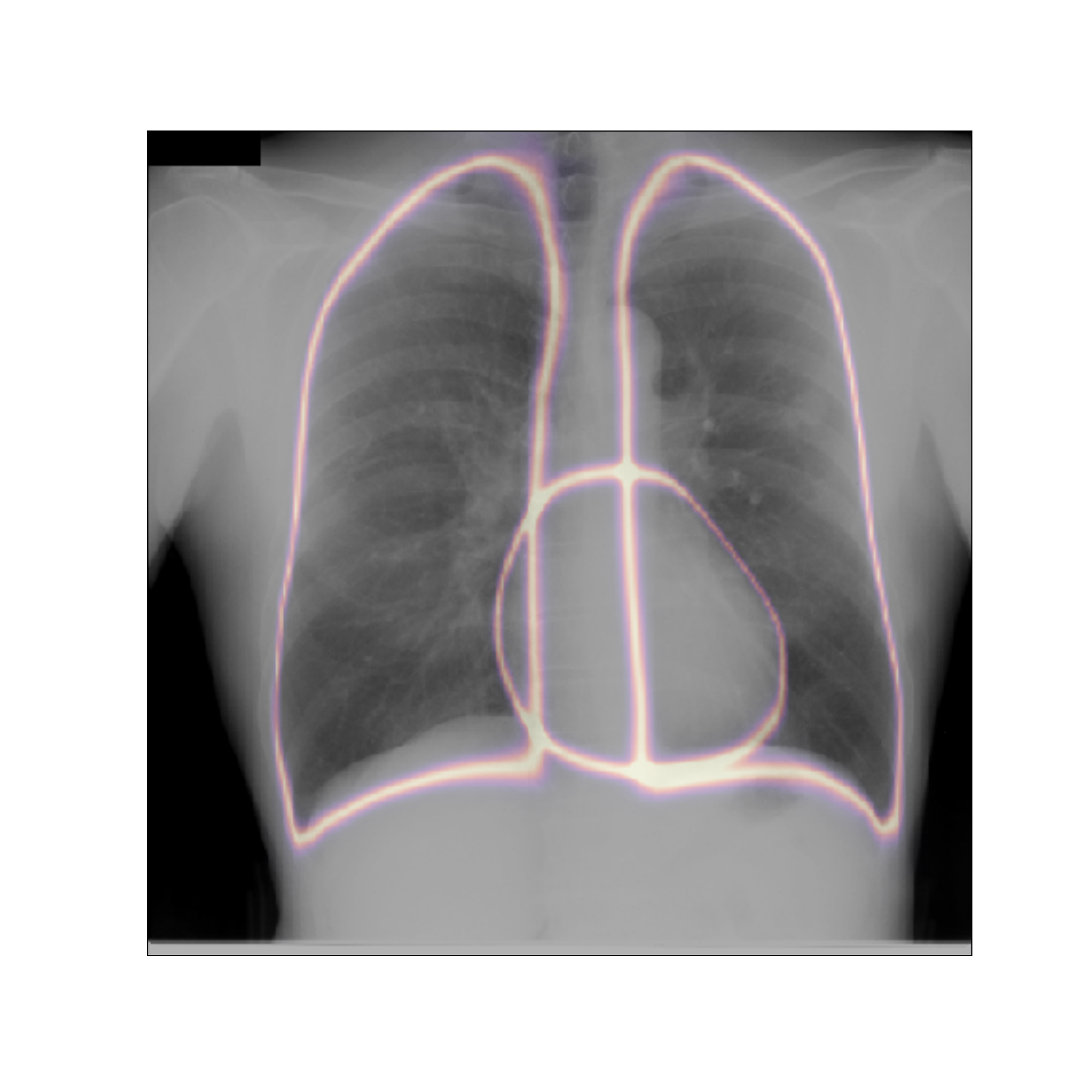}
 		\includegraphics[width=.24\textwidth]{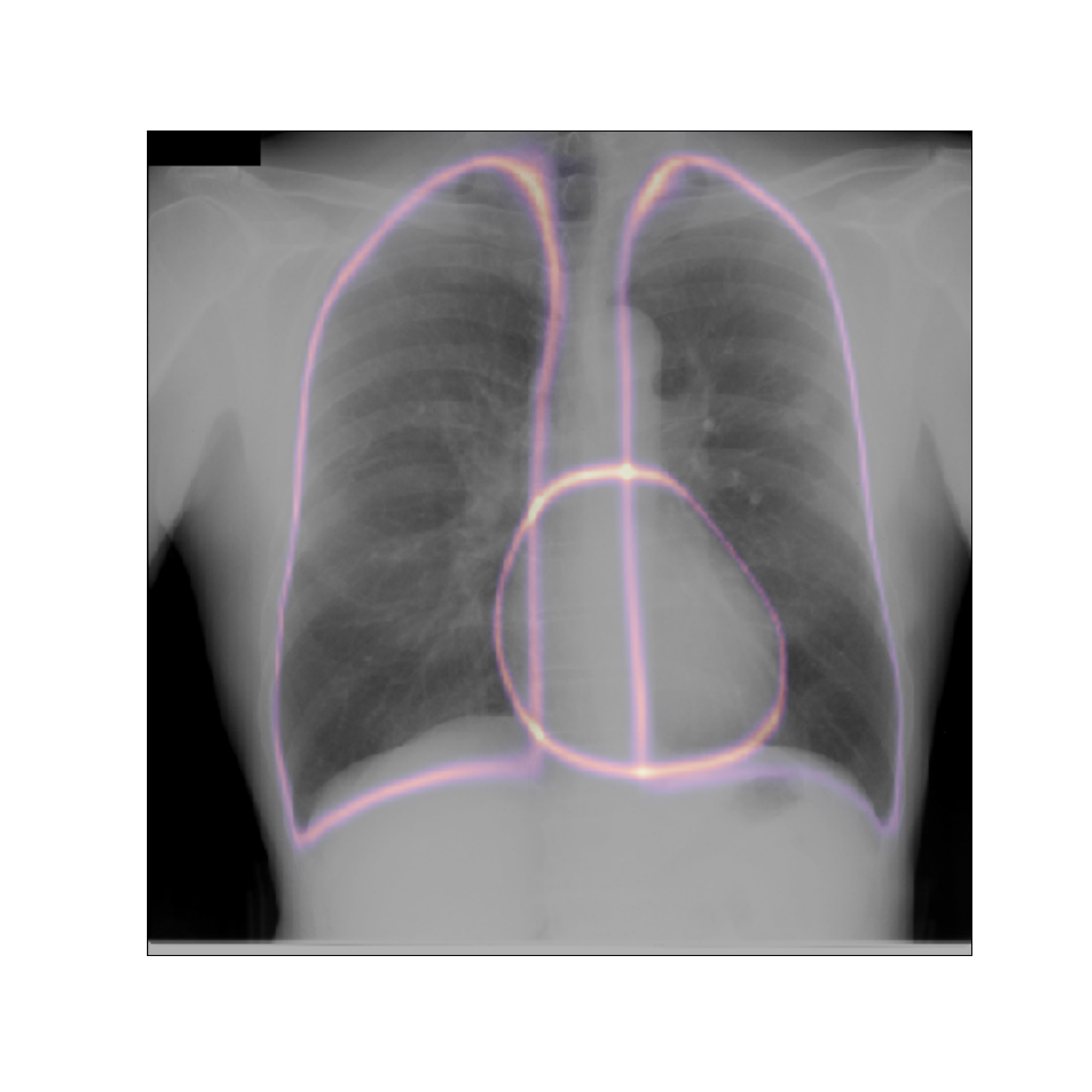}
 		\caption{JSRT}
 	\end{subfigure}

 	\begin{subfigure}[t]{\textwidth}
 		\centering
 		\includegraphics[width=.24\textwidth]{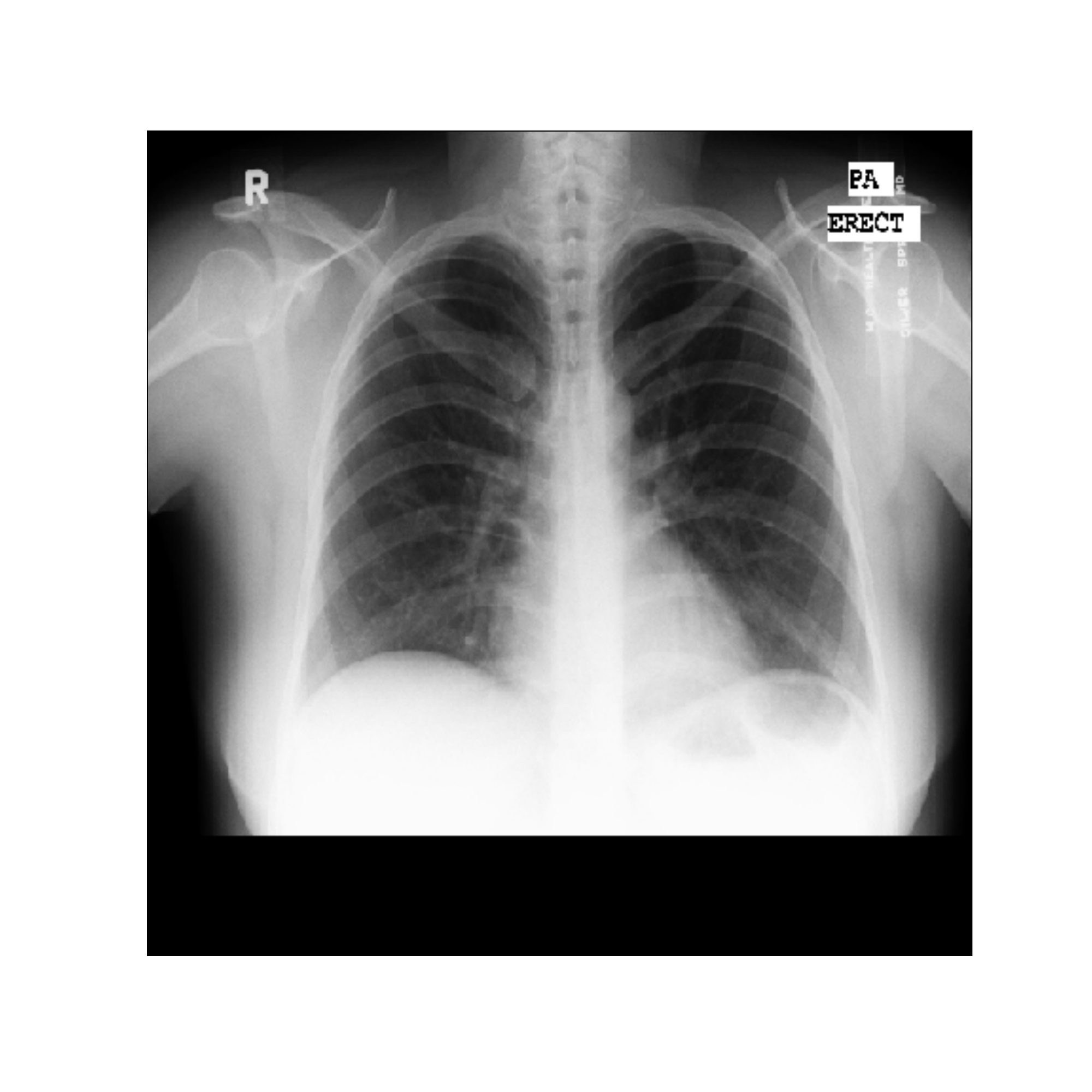}
 		\includegraphics[width=.24\textwidth]{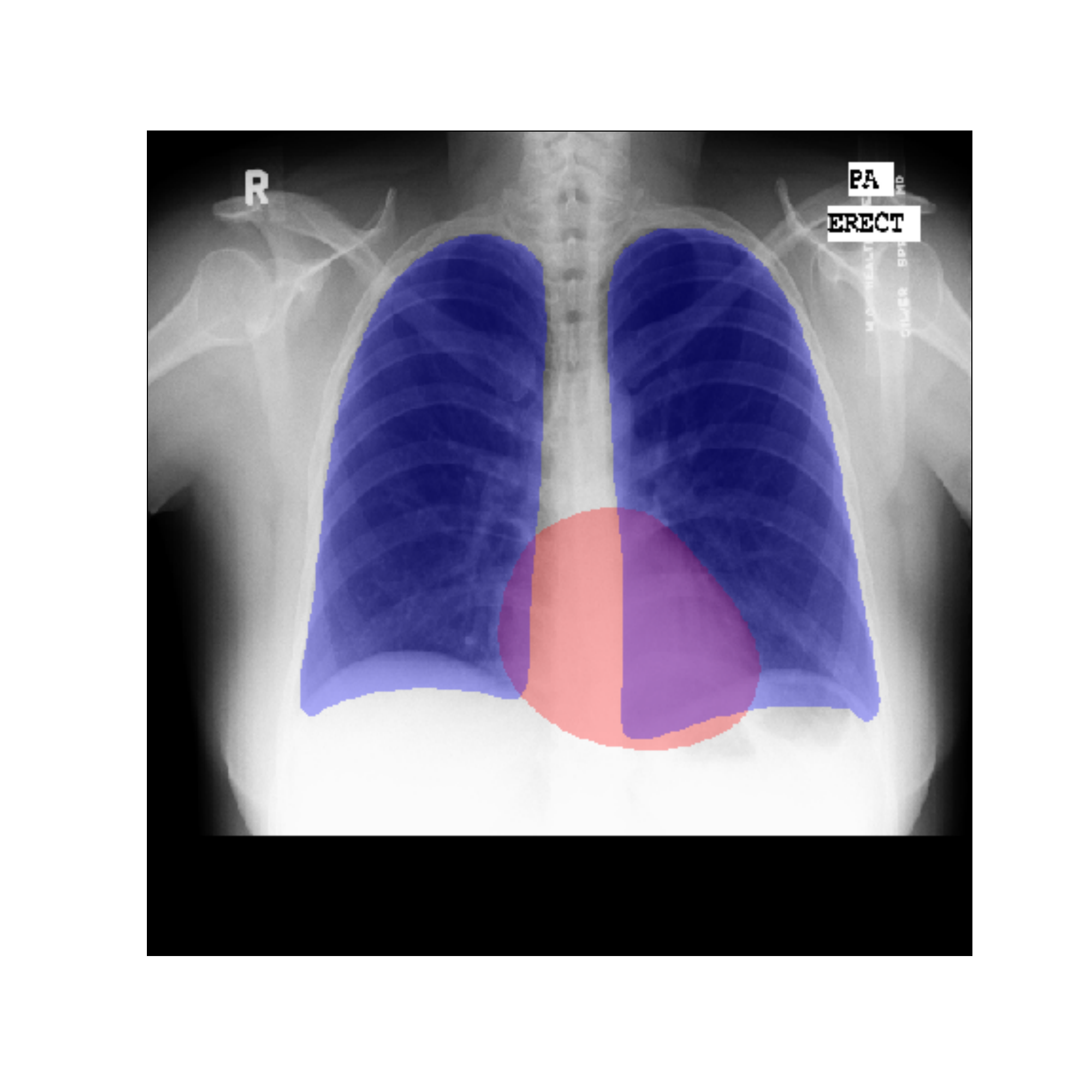}
 		\includegraphics[width=.24\textwidth]{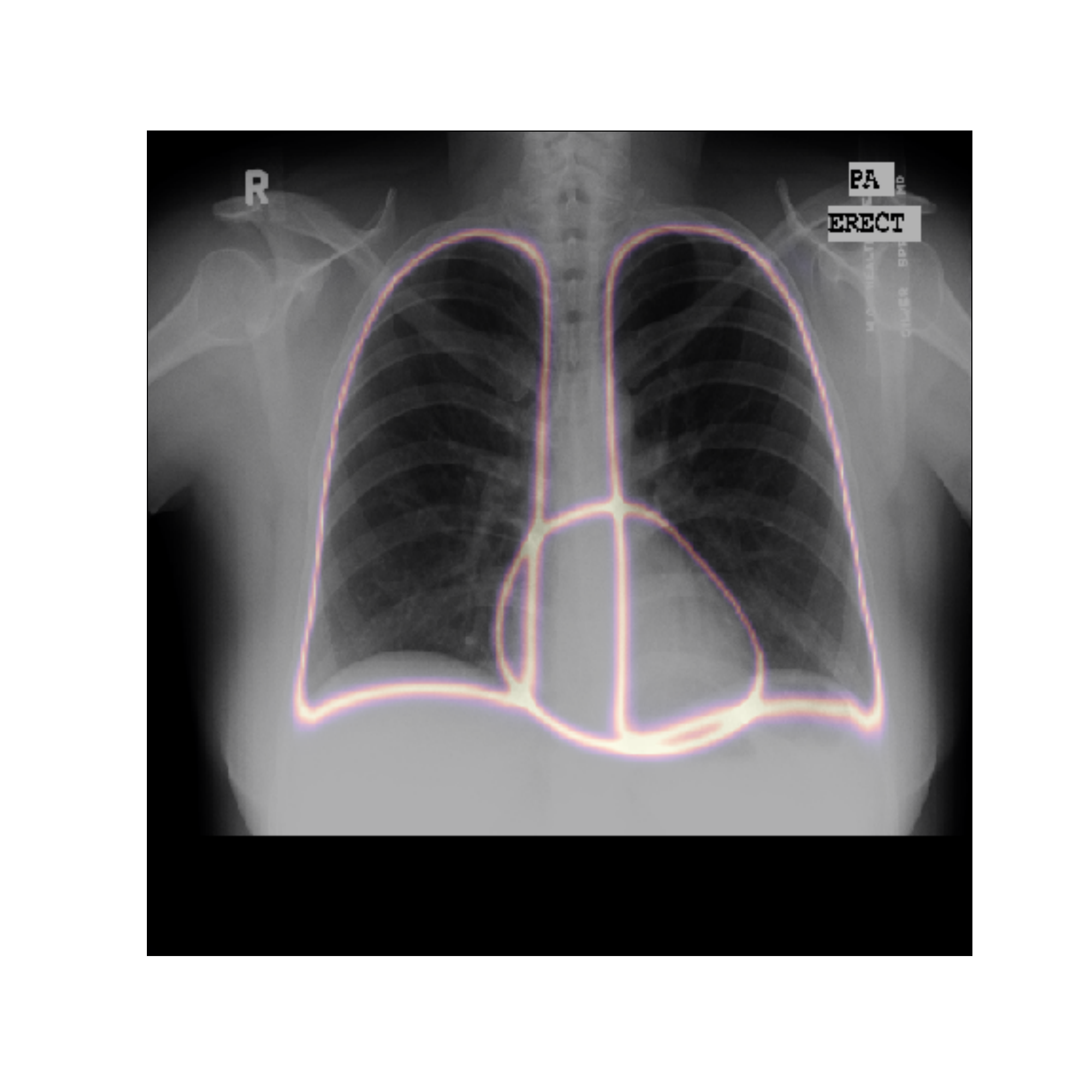}
 		\includegraphics[width=.24\textwidth]{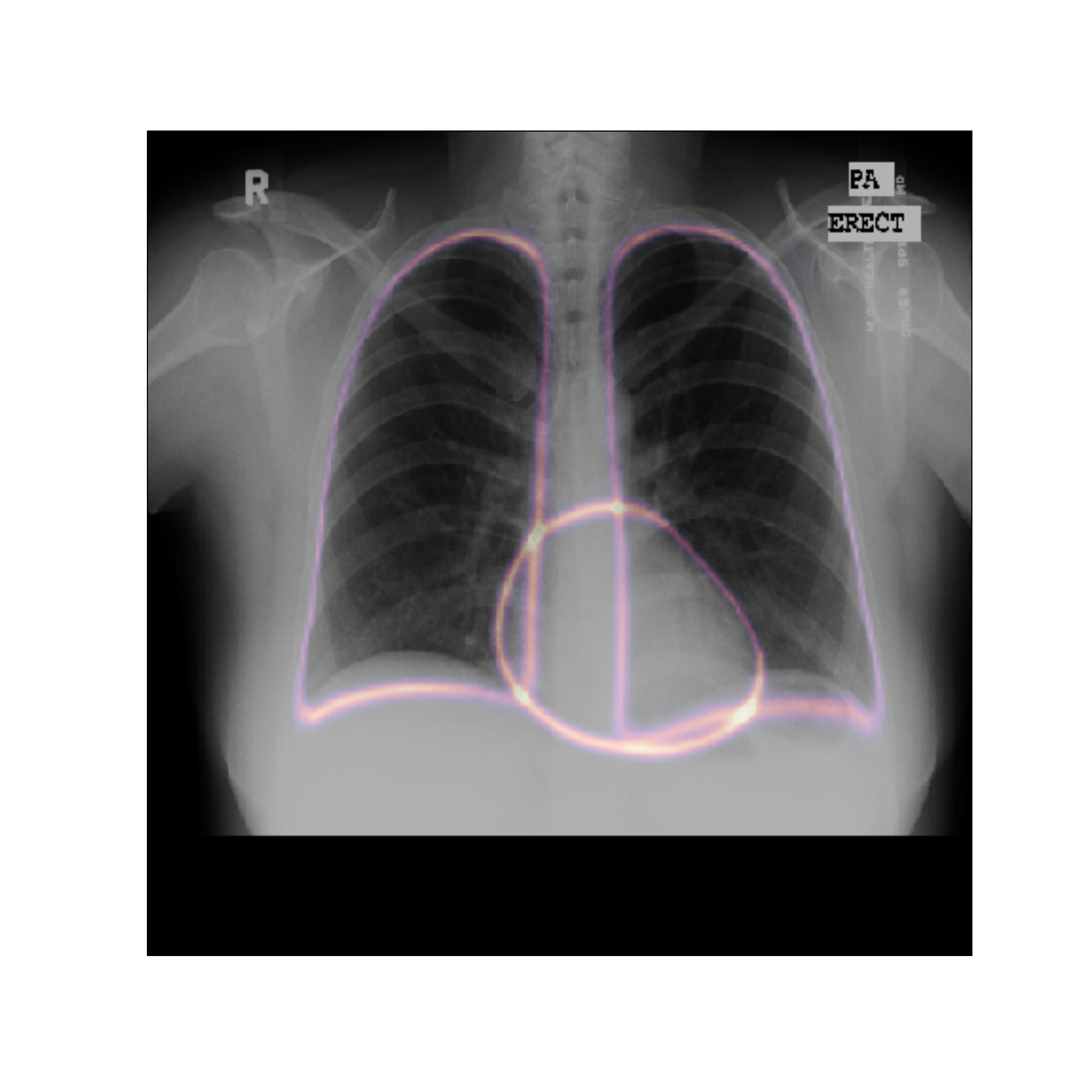}
 		\caption{Montgomery}
 	\end{subfigure}

    \begin{subfigure}[t]{\textwidth}
 		\centering
 		\includegraphics[width=.24\textwidth]{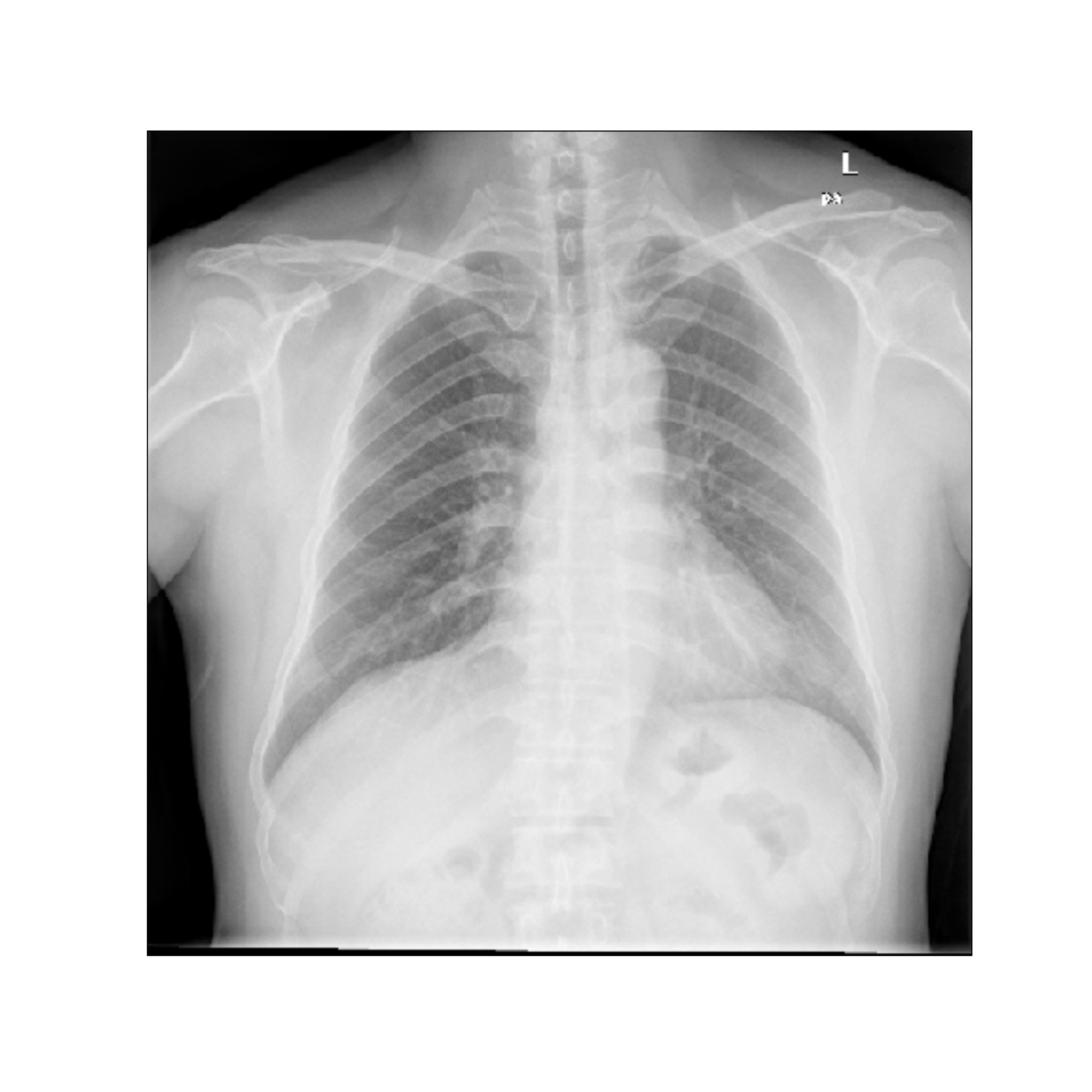}
 		\includegraphics[width=.24\textwidth]{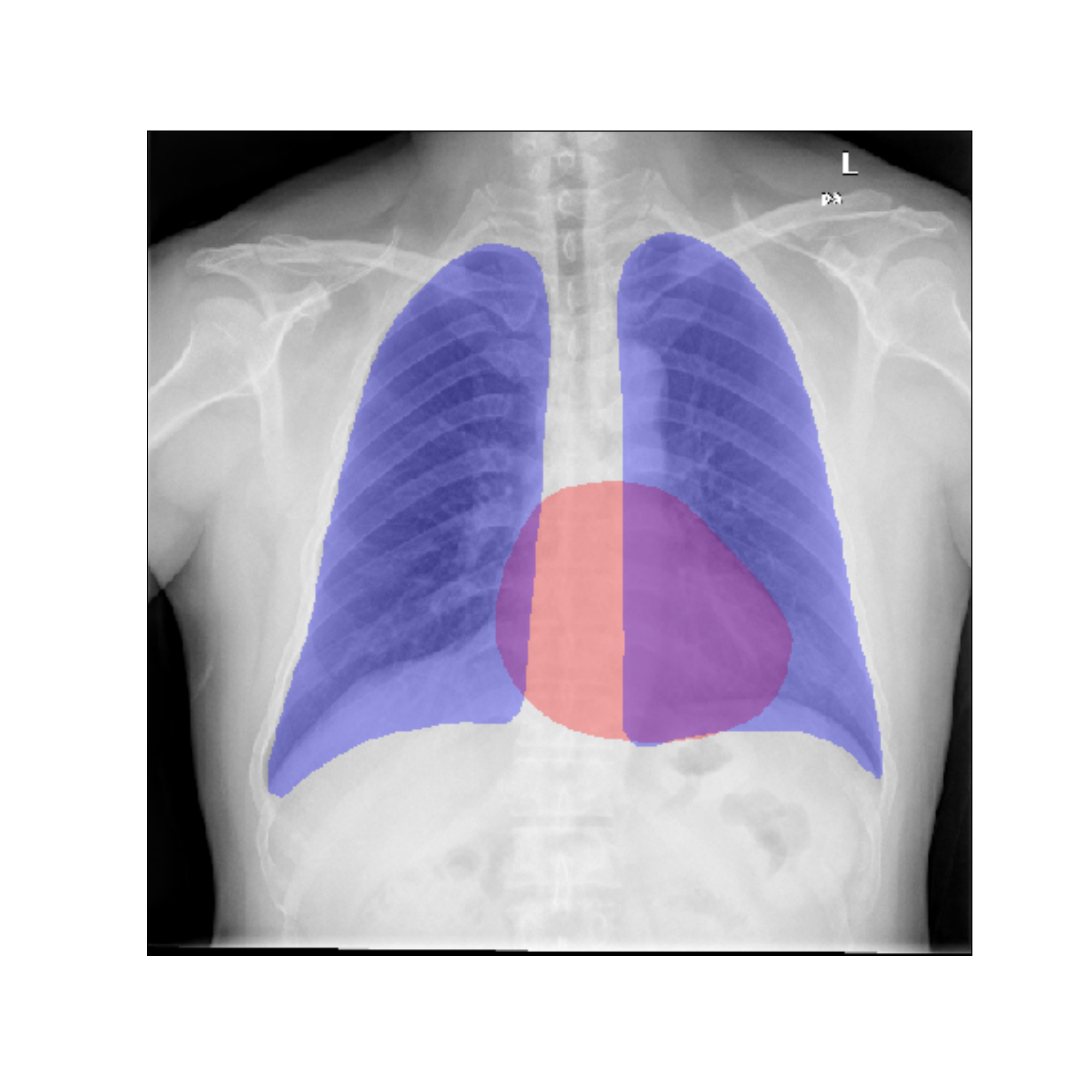}
 		\includegraphics[width=.24\textwidth]{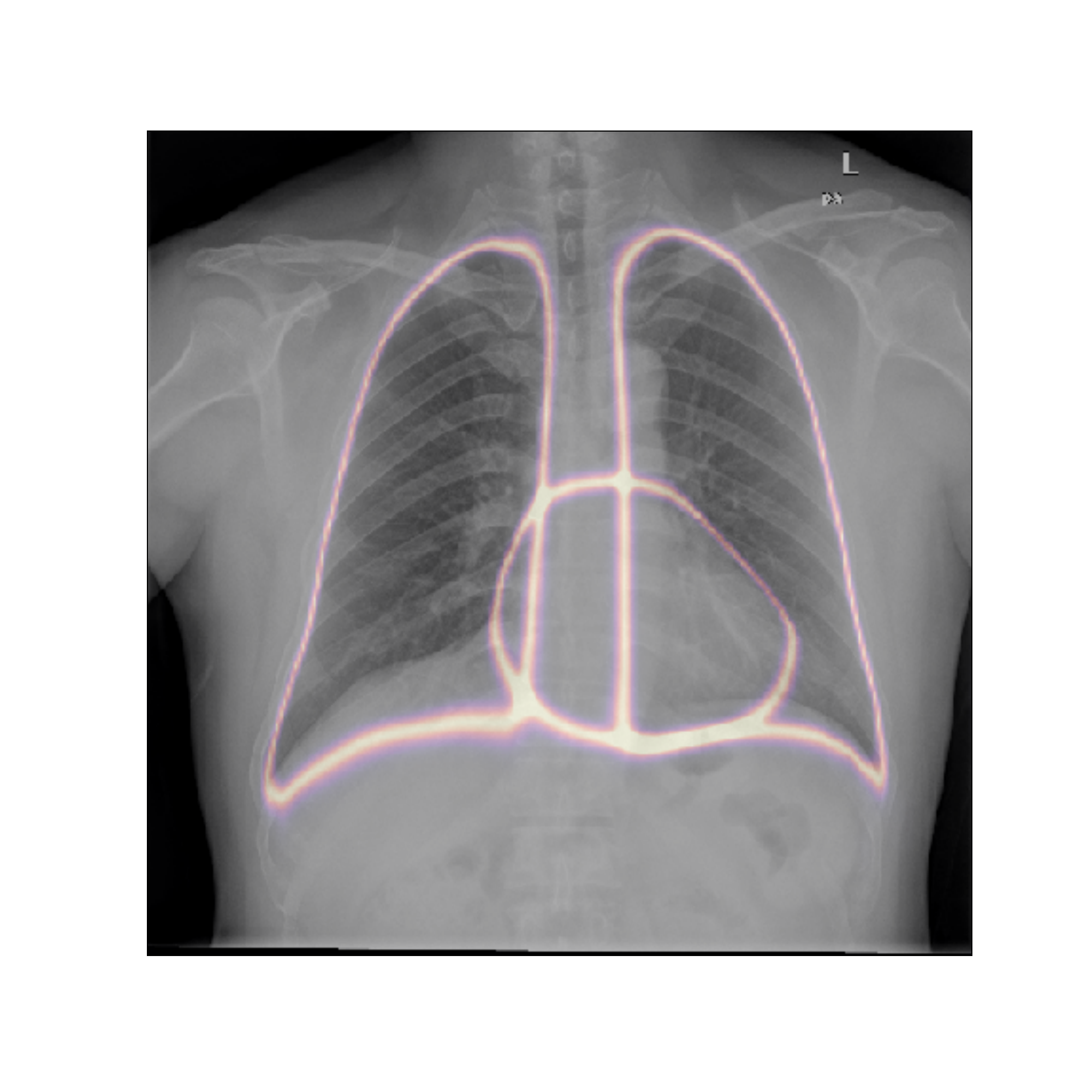}
 		\includegraphics[width=.24\textwidth]{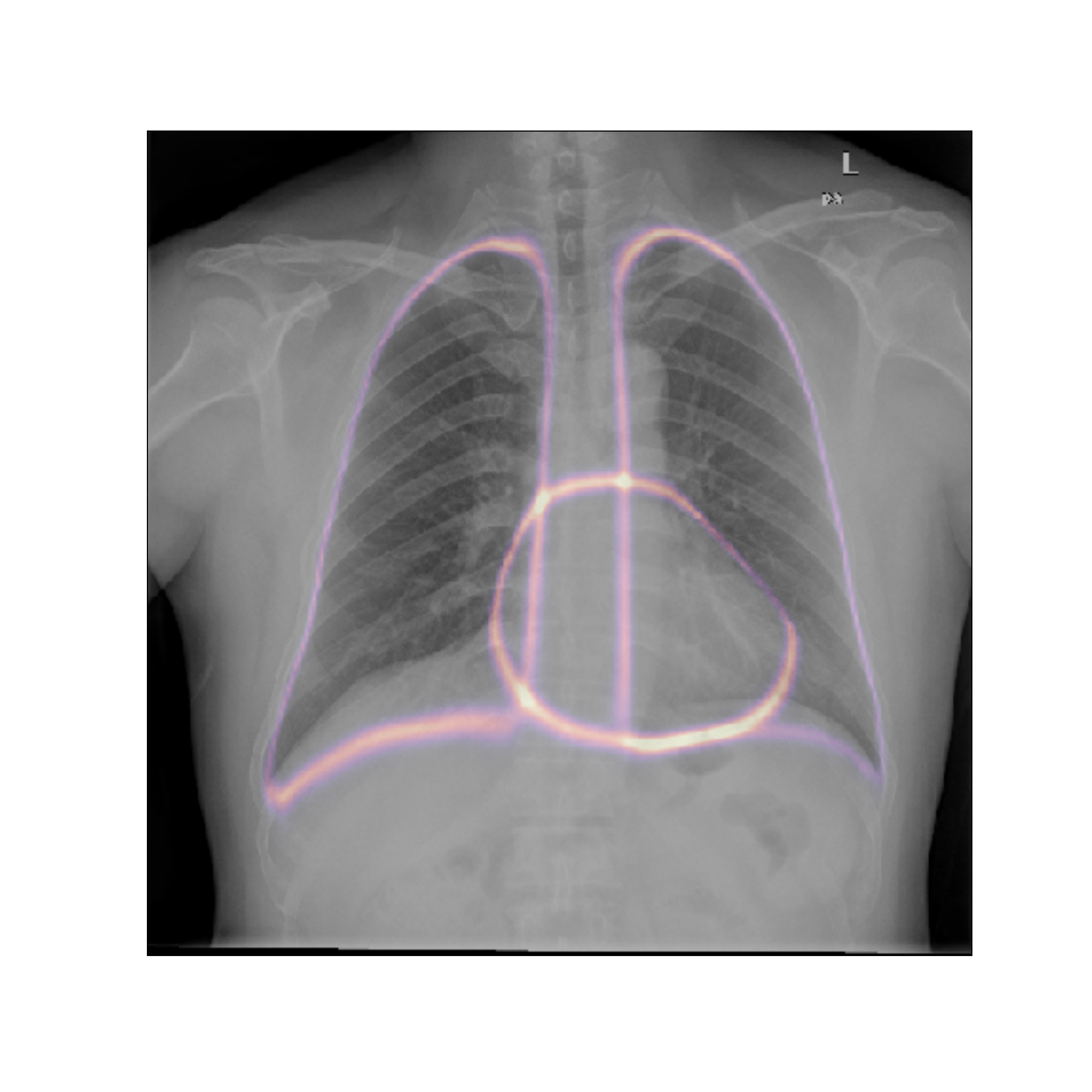}
 		\caption{Schenzhen}
 	\end{subfigure}
\caption{Examples of segmentation and uncertainty estimates for each of the test datasets (random examples are shown). Here, from left to right: original image, predicted segmentation mask, aleotoric and epistemic uncertainties, respectively.}\label{fig:results-example}
\end{figure*}

\paragraph{Segmentation Examples}
In Fig.~\ref{fig:results-example}, we visualized the examples of segmentation and the uncertainty estimates (both aleotoric and epistemic). The proposed method achieves accurate results demonstrating good segmentation performance in general. However, we note that our model does not predict the sharp corners of the lungs. Furthermore, from the epistemic uncertainty maps, it can be seen that our model is not confident in the bottom part of the lungs as they are typically very difficult to annotate since they can be hardly distinguished in the images. %Similar can also be observed for other regions, such as the ones where the spine overlaps with the heart boundary.

\section{Conclusion}
In this paper, we developed a novel approach for automatic segmentation of chest X-rays and assessment of CTR. Our approach is a modified FPN with ResNet50 backbone and MC dropout. In the extensive experimental evaluation, we found that the proposed configuration with instance normalization in the decoder yielded the best results compared to other investigated network configurations. Besides, it is worth to note that for the first time in CTR estimation realm, we proposed to assess it using Bayesian deep learning.

In this paper, we focused not only on developing state-of-the-art method for segmenting the chest X-rays, but also tackled the issue of annotation of these data and the availability of reliably annotated train and test data. As such, we proposed multiple new datasets that were annotated by radiologists and we plan to publicly release these data to facilitate further research.

Despite the advantages of our proposed method, this study has still some limitations. In particular, we did not experiment with training the models from scratch and used transfer learning. The second limitation of our study is that we did not compare our method to state-of-the-art unsupervised domain adaptation approaches~\cite{dong2018unsupervised, chen2018semantic,eslami2019image}. However, this would require re-implementation of previously presented methods as our annotations for all the test set differ from all the previously published techniques. We leave this limitation for the future work. Another important limitation of our study is that the annotators of the test data differ. In particular, one radiologist (radiologist A) annotated the train and the test sets derived from ChestXray14 dataset~\cite{Wang_2017_CVPR} and another radiologist (radiologist B) annotated the images from JSRT~\cite{shiraishi2000development}, Montgomery and Shenzhen datasets~\cite{jaeger2014two}. While we think that this particular limitation has insignificant impact onto our results, we still plan assess the inter-rater agreement between the annotators of the data.

To conclude, this paper introduced a novel, more challenging setting for segmenting organs in chest X-rays and proposed a Bayesian modification of FPN that allowed to estimate the CTR with the uncertainty bounds using MC-dropout. We think that the proposed approach has multiple applications in the clinical practice, as such, it could be useful for quantitative monitoring of CTR for patients with cardiomegaly in intensive care units. Another interesting application is the image quality assessment since our model is able to predict the aleotoric uncertainty for every test image. Finally, for the benefit of the community, we publicly release our dataset, implementation of our method and the pre-trained models at~\url{http://will.be.placed.after.review}.

{\small
\bibliographystyle{ieee}
\bibliography{egbib}
}

\end{document}